\pdfoutput=1
\documentclass[11pt,a4]{article}
\usepackage{graphicx}
\usepackage{epstopdf, epsfig}
\usepackage{amsmath}
\usepackage[numbers]{natbib} 
\usepackage{geometry}

\begin{document}


\title{Effect of spatial discretization of energy on detonation wave propagation}

\author{XiaoCheng Mi, Evgeny V. Timofeev, and Andrew J. Higgins\\[0.7em]
McGill University, Department of Mechanical Engineering, Montreal, QC, Canada}

\maketitle

\begin{abstract}
Detonation propagation in the limit of highly spatially discretized energy sources is investigated. The model of this problem begins with a medium consisting of a calorically perfect gas with a prescribed energy release per unit mass.  The energy release is collected into sheet-like sources that are now embedded in an inert gas that fills the spaces between them. The release of energy in the first sheet results in a planar blast wave that propagates to the next source, which is triggered after a prescribed delay, generating a new blast, and so forth.  The resulting wave dynamics as the front passes through hundreds of such sources is computationally simulated by numerically solving the governing one-dimensional Euler equations in the lab-fixed reference frame. Two different solvers are used:  one with a fixed uniform grid and the other using an unstructured, adaptively refined grid enabling the limit of highly concentrated, spatially discrete sources to be examined.  The two different solvers generate consistent results, agreeing within the accuracy of the measured wave speeds.  The average wave speed for each simulation is measured once the wave propagation has reached a quasi-periodic solution.  The effect of source delay time, source energy density, specific heat ratio, and the spatial discreteness of the sources on the wave speed is studied.  Sources fixed in the lab reference frame versus sources that convect with the flow are compared as well.  The average wave speed is compared to the ideal Chapman-Jouguet (CJ) speed of the equivalent homogenized media.  Velocities in excess of the CJ speed are found as the sources are made increasingly discrete, with the deviation above CJ being as great as $15\%$.  The deviation above the CJ value increases with decreasing values of specific heat ratio $\gamma$.  The total energy release, delay time, and whether the sources remain lab-fixed or are convected with the flow do not have a significant influence on the deviation of the average wave speed away from CJ.  A simple, \textit{ad hoc} analytic model is proposed to treat the case of zero delay time (i.e., source energy released at the shock front) that exhibits qualitative agreement with the computational solutions and may explain why the deviation from CJ increases with decreasing $\gamma$.  When the sources are sufficiently spread out so as to make the energy release of the media nearly continuous, the classic CJ solution is obtained for the average wave speed.  Such continuous waves can also be shown to have a time-averaged structure consistent with the classical ZND structure of a detonation.  In the limit of highly discrete sources, temporal averaging of the wave structure shows that the effective sonic surface does not correspond to an equilibrium state.  The average state of the flow leaving the wave in this case does eventually reach the equilibrium Hugoniot, but only after the effective sonic surface has been crossed. Thus, the super-CJ waves observed in the limit of highly discretized sources can be understood as weak detonations due to the non-equilibrium state at the effective sonic surface. These results have implications for the validity of the CJ criterion as applied to highly unstable detonations in gases and heterogeneous detonations in condensed phase and multi-phase media.
\end{abstract}

\section{Introduction}
\label{Sec_Intro}
Detonation waves are combustion waves that move at supersonic speed with respect to the energetic medium through which they propagate. For a wave propagating at constant velocity, the steady conservation laws of mass, momentum, and energy can be applied to a control volume that encloses and moves with the wave.  Along with an equation of state, an additional criterion is required to close the set of governing equations and solve for the wave velocity and thermodynamic state at the exit of the wave.  The condition of sonic flow (with respect to the wave-fixed frame) was proposed by \citet{Chapman1899} and \citet{Jouguet1905}, and this state can be shown to correspond to the minimum possible speed of a compressive wave with equilibrium products that satisfies the steady conservation laws.  The Chapman-Jouguet (CJ) criterion can be justified on the grounds that subsonic flow at the exit of the control volume would allow rarefaction waves overtaking the detonation from downstream to weaken and decelerate the wave.  Thus, a detonation overdriven by a piston from behind with subsonic flow at the end of the reaction zone (a so-called strong detonation) would eventually relax to the CJ solution when the driving piston is stopped.  A supersonic state exiting a self-sustained wave (a weak detonation) might be possible under the scenario of complex chemical kinetics wherein the reaction zone is described by an exothermic phase, which brings the non-equilibrium flow to sonic, followed by an endothermic phase.  In these so-called pathological detonations, rarely observed in experiments, the CJ criterion is replaced by an equivalent generalized CJ criterion that applies to the non-equilibrium flow in the reaction zone \citep{Eyring1949,WoodKirkwood1954,HigginsChapter2}. The classical CJ criterion has been remarkably successful in predicting detonation velocity, with experimentally observed detonation waves in gases usually observed to propagate within $1\%$ of the equilibrium CJ speed in nearly all gaseous mixtures.  The CJ criterion has had similar success in condensed phase energetic materials (i.e., solid and liquid high explosives), although uncertainty in the equation of state of condensed phase materials at extremely high pressure means that these calculations are not usually conducted from first principles and must rely on empirical correlations (often using the measured detonation velocity of the explosive itself as an input parameter).

The success of the CJ criterion is all the more remarkable given that all known detonations in gases are unstable, dominated by a transient and multidimensional cellular structure. The fact that detonations, governed by activated chemical reactions (i.e., Arrhenius kinetics), are unstable to perturbations was first demonstrated by \citet{Erpenbeck1964}.   \citet{Short1998} proved that a detonation with a realistic value of heat release and activation energy would be unconditionally unstable in two-dimensions, resulting in a cellular structure.  The experimental evidence of detonation cells was first noted by \citet{Denisov1961}, and these detonation cells are now recognized as being essential for the propagation of detonations in high activation energy mixtures (e.g., hydrocarbon fuels in air).  In mixtures with high activation energy (e.g., methane/oxygen/diluent \citep{Strehlow1968, Strehlow1969}) or with gases featuring a large number of molecular degrees of freedom, resulting in low values of heat capacity ratio (e.g., halocarbons \citep{Vndermeiren1988,Lefebvre1993}), the detonation cellular structure becomes highly irregular, with explosive bursts of fine cellular structure and regions of apparently no cells at all.  In such mixtures, deviations in local detonation front velocity can be as much as $50\%$ below and $100\%$ above the mean velocity.  Despite these complexities, the CJ criterion is still remarkably accurate in predicting the overall average speed of the detonation front.  In liquid explosives and some solid explosives such as TNT, similar cellular instabilities are also noted \citep{Fickett2000,Lee2002}.  Other explosives, such as those typically used in commercial blasting, have a highly heterogeneous structure due to the coarse grains of energetic material used, with the scale of the heterogeneity being on the order of the detonation front thickness.  Again, in these cases, the CJ criterion appears to be remarkably successful at predicting the detonation velocity of effectively infinite diameter charges (although the earlier caveat about equation of state still applies).

The seemingly paradoxical situation, in which the steady, one-dimensional CJ criterion appears successful in describing the propagation of what is now known to be a highly complex wave structure, was well articulated by Oppenheim shortly after the discovery of cellular structure in gases:\begin{quote}
The detonation may form an essentially non-steady, non-uniform regime so that in order to explain its precise nature, multi-dimensional effects in space as well as its irregular behavior in time have to be taken into account. In fact, in view of this evidence one should express amazement that the one dimensional steady flow theory was so successful in rationalizing so many experimental observations \citep{Oppenheim1961}.
\end{quote}
A half century later, this puzzle has not been completely resolved.  The usual argument that is made is that, if a sufficiently large control volume is used to enclose the wave, then the steady, one-dimensional control volume analysis must still apply on average, even though the wave itself is multidimensional and unsteady.  This reasoning is not entirely satisfactory, however.  For one thing, it is not clear where to place the exit plane of the control volume \textit{a priori} (too close to the initial shock front will locate the exit plane in a non-equilibrium flow, while too far downstream places the exit plane in an unsteady rarefaction, the Taylor wave, where the steady conservation laws do not apply).  In addition, for an unsteady flow, the concept of sonic flow is not particularly relevant (i.e., which reference frame should be used in comparing the flow velocity to the local speed of sound?).  For unsteady detonation propagation in one-dimension, the sonic surface is more accurately replaced by the limiting separatrix of characteristic waves propagating from behind the detonation toward the shock front, with characteristics in front of the separatrix able to influence the shock and those behind not able to reach the shock \citep{KasimovStewart2004, NgChapter3}. The applicability of the steady, one-dimensional analysis to a two-dimensional, moderately unstable detonation wave was explored computationally by \citet{Radulescu2007JFM} via spatial and temporal averaging of the unsteady structure in a steady reference frame following the wave. In their study, the effect of the interplay between the chemical exothermicity and hydrodynamic fluctuations on the location of the effective sonic surface in a two-dimensional, unsteady detonation structure was clearly illustrated by this resulting one-dimensional averaged wave profile \citep{Radulescu2007JFM}.

\begin{figure}[h]
\center
\includegraphics[width=0.9\textwidth]{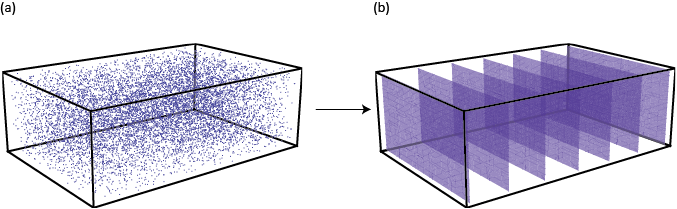}
\caption{Schematic of an explosive medium with ($a$) source energy uniformly distributed and with ($b$) source energy collected into planar sheets.}
\label{Fig1}
\end{figure}

In this paper, we intend to put the CJ criterion to a rigorous test by examining detonation in a media in which the energy release is condensed into spatially discrete layers separated by regions of inert gas (as illustrated in figure~\ref{Fig1}).  The energetic layers will release their energy when activated by a passing shock front after a fixed delay time.  The new blast wave generated by the energy release becomes the mechanism to initiate additional sources, such that the wave propagates via a sequence of ``sympathetic detonations.'' The spatial scale of the sources and their spacing will be varied, examining the behavior of the wave propagation in this system from continuous energy release (i.e., no discretization) to highly concentrated sources in the limit of $\delta$-function-like sources of energy separated by layers of inert gas.  The overall energy release will be maintained as constant, so that the average wave velocity can always be compared to the equivalent CJ speed of the homogenized media.  This work builds upon similar research examining flame propagation in discretized media, in which point-like sources release heat to diffuse outward and trigger subsequent sources \citep{GoroshinLeeShoshin1998,Beck2003,Tang2009CTM,Goroshin2011PRE,Tang2011PRE}. A wave propagation mechanism by which an energetic source generates a blast that, in turn, can initiate the next source was proposed by \citet{Stewart1991} in modeling the response of propellant beds comprised of explosive grains to strong shock stimuli. They outlined a ``theory of discrete interactions'' in which initiation of subsequent sources would be described by a nonlinear recursion relation. In spray detonations with millimeter-sized droplets of fuel, a similar propagation mechanism, i.e., blast waves generated by each individual droplet releasing its energy, merging with and reinforcing the leading shock wave, was experimentally identified by \citet{Dabora1969} and theoretically described \citet{Pierce1973}. Propagation of detonation in media with a sinusoidal variation in properties was explored computationally by \citet{Morano2002} in one dimension and extended to two dimensions by \citet{Li2014CS}. In these studies, no significant deviation greater than $2\%$ away from CJ was observed when the detonation propagated without losses present.  \citet{Li2014CS} did observe enhanced wave speeds when the detonation propagated in a layer with inhomogenities and yielding confinement (in comparison to a detonation propagating in a homogeneous layer with the same yielding confinement).  The enhanced wave speeds were attributed to accelerated reaction rates resulting from the shock interactions caused by the inhomogeneities.  Detonation propagation in a media described by the Euler equations with $\delta$-function-like sources has not been previously published.

The plan for this paper is as follows.  In \S~\ref{Sec_Problem_Statement}, the problem statement and parameters will be defined.  Section~\ref{Sec_Numerical_Method} describes the numerical methodology used to solve the governing Euler equations.  Section~\ref{Sec_Results} presents the results of a study where each of the model parameters was systematically varied and compared to the ideal CJ detonation velocity of the equivalent homogenous media.  Section~\ref{Sec_Analysis} analyzes the results by temporally averaging select simulations to provide a quasi-steady one-dimensional description of the transient waves.  The results are discussed in \S~\ref{Sec_Discussion} and summarized in the Conclusions (\S~\ref{Sec_Conclusion}). An Appendix presents an \textit{ad hoc}, or heuristic, model that attempts to construct an analytic solution to this problem for the case of zero delay in triggering the sources. 
 
\section{Problem statement}
\label{Sec_Problem_Statement}
This study considers a calorically perfect gas (i.e., fixed ratio of specific heats $\gamma$) that has the potential to release energy with a heating value of $\widetilde{Q}$ ($\mathrm{J}/\mathrm{kg}$). The tilde ``$\sim$'' denotes a dimensional quantity. The flow variables, density, pressure, and particle velocity, are non-dimensionalized with reference to the initial state ahead of the leading shock, i.e., $\rho=\widetilde{\rho}/\widetilde{\rho}_0$, $p=\widetilde{p}/\widetilde{p}_0$, and $u=\widetilde{u}/\sqrt{\widetilde{p}_0/\widetilde{\rho}_0}$, respectively, and the space coordinate $x$ by the spacing between two adjacent sources $\widetilde{L}$, i.e., $x=\widetilde{x}/\widetilde{L}$. The subscript ``$0$'' indicates the initial state of the uniform reactive medium. The heat release $\widetilde{Q}$ is non-dimensionalized as $Q=\widetilde{Q}/\left( \widetilde{p}_0 / \widetilde{\rho}_0 \right)$.  A detonation wave propagating through this uniform medium is expected to move at the non-dimensional CJ speed, given by,
\begin{equation}
V_{\mathrm{CJ}} = M_{\mathrm{CJ}} c_0 = \sqrt{\frac{\gamma^{2}-1}{\gamma}Q+\sqrt{\left( \frac{\gamma^{2}-1}{\gamma}Q+1 \right)^{2}-1}+1} \cdot \sqrt{\gamma}
\label{Eq1}
\end{equation}
where $c_0$ denotes the non-dimensionalized initial speed of sound and equals $\sqrt{\gamma}$. Equation~\ref{Eq1} is the classical CJ detonation solution, and is used to compare the resulting propagation speeds from the discrete source simulations with that of the steady detonation in the equivalent homogeneous media. All the simulation results of wave speeds reported in this paper are normalized by the corresponding $V_{\mathrm{CJ}}$.

To examine the effect of spatial discretization, the energy is collected in regions with width $\widetilde{W}$ (source width) that are spaced a distance $\widetilde{L}$ apart. The total energy release of the medium remains fixed at $Q$, so the energy in each concentrated source is $Q\widetilde{L}/\widetilde{W}$. The discreteness of the system is described by a parameter $\Gamma=\widetilde{W}/\widetilde{L}$. As $\Gamma \to 0$, the sources of energy become highly spatially concentrated. In the limit of $\Gamma \to 1$, the energy release becomes continuous through the medium. A source is assumed to be triggered by the passage of a rightward-propagating shock wave past the right edge of the source, independent of the strength of the shock, after a time delay $t_\mathrm{d}$. This time delay is reported as a non-dimensional value $\tau=t_\mathrm{d} V_{\mathrm{CJ}}$ , where the time required for the CJ detonation in the equivalent homogenous system has been used to non-dimensionalize the delay time. Note that the non-dimensionalized spacing between two adjacent sources is unity.

Two different scenarios of sources are considered in this study, fixed and convected. For fixed sources, the energy source is considered to be independent of the inert, gaseous media and is not convected along with it. Conceptually, this would correspond to a fine mesh of explosive wire in a tube filled with inert gas, wherein the mesh spans the tube cross-sectional area and is attached to the tube wall.  In this case, the shock-accelerated, inert gas is free to flow (without resistance) around the sources of energy, which remain fixed until they release their energy. In the second scenario (convected sources), the sources are assumed to be embedded in the inert, gaseous media and convect along with it. This would correspond to a layer of explosive gas or explosive dust, suspended between sections of inert gas. As the successive blast waves propagate through the media, the sources in this scenario are compressed and convected along with the post-shocked inert gas. These two scenarios are illustrated schematically in figure~\ref{Fig2}.

There is no explicit source term in the governing Euler equations. After the delay time has elapsed, the simulation is paused, and the energy that is instantaneously released by the source is added in the form of pressure to the volume occupied by the source at that moment. Maintaining the overall amount of heat release $Q$, the pressure increase in the region of a source can be calculated as follows
\begin{equation}
\mathrm{\Delta} p = \frac{Q\left( \gamma-1 \right)}{\Gamma}
\label{Eq3}
\end{equation}
The simulation is resumed after $\mathrm{\Delta} p$ is added. It is important to note that, in the scenario of convected sources, the actual source width $\widetilde{W}$ at the instant of energy release is smaller than its initial value, due to the compression of the media in which the source energy is stored and the particle motion induced by the passage of the shock wave. The numerical implementation of delay time and convected source tracking and energy addition is described in \S~\ref{Sec_Numerical_Method}.

Except for the instants in which the sources release energy, the system is described by the one-dimensional, nonreactive Euler equations in the lab-fixed reference frame,
\begin{equation}
\frac{\partial \bf{U}}{\partial t} + \frac{\partial \bf{F}\left( \bf{U} \right)}{\partial x} = 0
\label{Eq4}
\end{equation}
where the conserved variable $\bf{U}$ and the convective flux $\bf{F}$ are, respectively,
\begin{equation}
 \bf{U}=
 \begin{pmatrix}
 \rho \\
\rho u \\
\rho e
 \end{pmatrix}
~\qquad \bf{F}\left( \bf{U} \right) =
 \begin{pmatrix}
 \rho u \\
\rho u^2+p \\
(\rho e+p)u 
 \end{pmatrix}
\label{Eq5}
\end{equation}
In the above equations, $e$ is the non-dimensional specific total energy. The numerical solution methodology is described in the next section.

\section{Numerical method}
\label{Sec_Numerical_Method}
Two independently-written, one-dimensional, finite-volume Euler codes were used in this study. One of them was based on a fixed (non-adaptive) uniform grid. This solver, however, encountered computer memory limitations as the sources were progressively made more discrete, due to the requirement for a minimum number of computational cells within each source to properly resolve all wave phenomena induced by energy release  (see discussion below). For this reason, a second code using adaptive mesh refinement was used to both extend the results into the limit of highly spatially discrete sources and verify the results of the fixed-grid code. Both codes solve the one-dimensional Euler equations using the MUSCL-Hancock TVD node-centered Godunov-type finite-volume scheme \citep{Toro2009} with an exact Riemann solver and the van Leer non-smooth slope limiter. The scheme is of second order of accuracy in space and time on smooth solutions.

In the adaptive code, the background (initial) grid is also uniform. Grid adaptation is performed at each time step via hierarchical \textit{h}-refinement. Each refinement level reduces the local grid step by two times. The number of refinement levels is determined from the requirement of having the smallest grid step corresponding to the desired number of grid nodes within the energy source (see discussion below). The refinement/coarsening sensor is based on normalized second derivatives of density and pressure. Prior to energy release via pressure modification for a source, the grid is refined in the vicinity of the source, regardless of the sensor values in the region, so that the energy release and subsequent induced wave motion would be always resolved on the finest mesh. The data structure and adaptation procedure of the adaptive code were adopted, with suitable simplification, from the 2-D unstructured Euler code described by \citet{Satio2001} to a 1-D code. Preliminary numerical trials demonstrated that simple linear solution interpolation used in the code \citep{Satio2001} when inserting a new node led to significant (of the order of $5\%$) error in wave speeds when long distance propagation typical for the problem under consideration was simulated. For this reason, node insertion and deletion procedures were modified to be fully conservative, i.e., the conservation of mass, momentum, and energy were strictly enforced when assigning gasdynamic parameters at a newly inserted node or removing a node. 

\begin{figure}
\centerline{\includegraphics[width=0.9\textwidth]{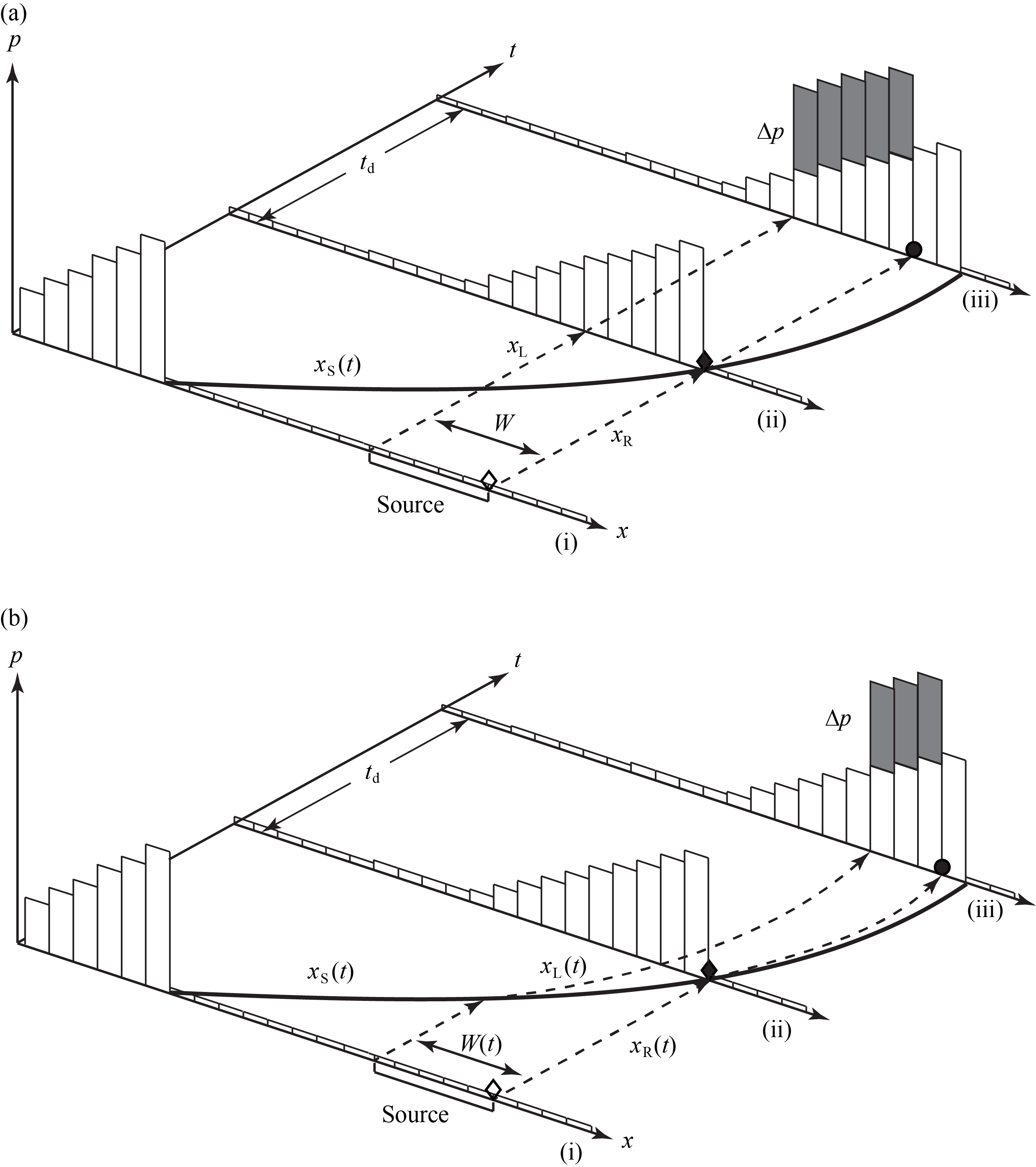}}
		\caption{Schematic of the numerical implementation of the energy release process and source tracking algorithm for ($a$) lab-fixed and ($b$) convected discrete sources at three different moments: (i) before the arrival of the leading shock wave, (ii) the source being shocked marking the onset of delay period, and (iii) the source releasing energy.}
	\label{Fig2}
\end{figure}

In both fixed and adaptive mesh refinement codes, the energy release of each discrete source after a finite delay period is implemented as follows. The trajectory of the leading shock front, $x_\mathrm{s}(t)$, is tracked in the simulations by finding the location where $p$ first increases to $1.01$ from its upstream initial state $p_0=1$ at each time step.  A discrete source is considered as being shocked once the leading shock passes the right edge of the space occupied by this source, i.e., the pressure at right edge of the source ($\left. p\right \vert_{x = x_\mathrm{R}}$) becomes greater than $p_0$. A variable, $t_\mathrm{i}$, with its initial value $0$, is then introduced to measure the time elapsed since the source has been shocked. The algorithm of delay time tracking is formulated as follows,
\begin{equation}
\frac{\mathrm{d} t_\mathrm{i}}{\mathrm{d} t}=\left\{
\begin{array}{c l}      
    0 \;\;\;\; & \left. p\right \vert_{x = x_\mathrm{R}} = p_0 \\[0.5em]
    1 \;\;\;\; & \left. p\right \vert_{x = x_\mathrm{R}} > p_0 
\end{array}\right.
\label{eq_delay}
\end{equation}
Once $t_\mathrm{i}$ reaches $t_\mathrm{d}$, the simulation is stopped before advancing to the next time step, and the pressure within the space occupied by this triggered source, i.e., between the left and right edges of this source ($x_\mathrm{L} \leq x	\leq x_\mathrm{R}$), is increased by $\mathrm{\Delta}p$,
\begin{equation}
\left. p\right \vert_{x_\mathrm{L} \leq x	\leq x_\mathrm{R}}  = \left. p\right \vert_{x_\mathrm{L}	\leq x \leq x_\mathrm{R}} + \mathrm{\Delta}p \;\;\;\;\mathrm{at} \;\;\;\; t_\mathrm{i} = t_\mathrm{d}
\label{eq_add_p}
\end{equation}
Unlike fixed sources, the location and volume occupied by a convected source change with time. The trajectories of the fluid elements marking the right and left edges of each convected source, denoted as $x_\mathrm{R} (t)$ and $x_\mathrm{L} (t)$, respectively, are tracked via numerical integration of the corresponding particle paths using a Lagrangian approach,
\begin{equation}
\begin{array}{c l}      
    x_\mathrm{R}^{t+\mathrm{\Delta}t} =&  x_\mathrm{R}^{t}+\mathrm{\Delta}t \cdot u_\mathrm{R}^{t} \\[0.3em]
    x_\mathrm{L}^{t+\mathrm{\Delta}t} =&  x_\mathrm{L}^{t}+\mathrm{\Delta}t \cdot u_\mathrm{L}^{t}
\end{array}
\label{Eq7}
\end{equation}
With this source tracking algorithm, the code is able to add pressure into the space occupied by a convected source at the moment when it releases energy after the delay period. The numerical implementation of energy release and source tracking algorithm for lab-fixed and convected sources is illustrated in figure~\ref{Fig2}.
As the numerical resolution is increased from $50$ to $200$ computational cells within a discrete source, the simulation results did not show any significant difference. The results reported in this paper were all performed with a numerical resolution of $100$ computational cells per source. 

The length of the entire simulated problem is over $100$ spatial units, i.e., containing more than $100$ discrete sources. In order to make the simulation time-efficient, the computation, at every time step, is only performed in a window enclosing the leading wave complex in a laboratory-fixed reference frame, instead of the entire simulation domain. A minimum eight-unit-wide window (i.e., containing $8$ sources) is necessary to capture all of the dynamics contributing to the propagation of the leading wave front for the longest delay times considered in this study.  Once the leading front reaches the end of this computational window, the window frame (i.e., left and right boundaries) advanced by one spatial unit while maintaining the same width of the domain. A transparent boundary condition is applied on both boundaries of the computational window \citep{Toro2009}. In order to rule out any possible influence of the boundary conditions, a $10$-unit-wide window was used in all the simulations reported below.

\section{Results}
\label{Sec_Results}
\begin{figure}
\centerline{\includegraphics[width=1.0\textwidth]{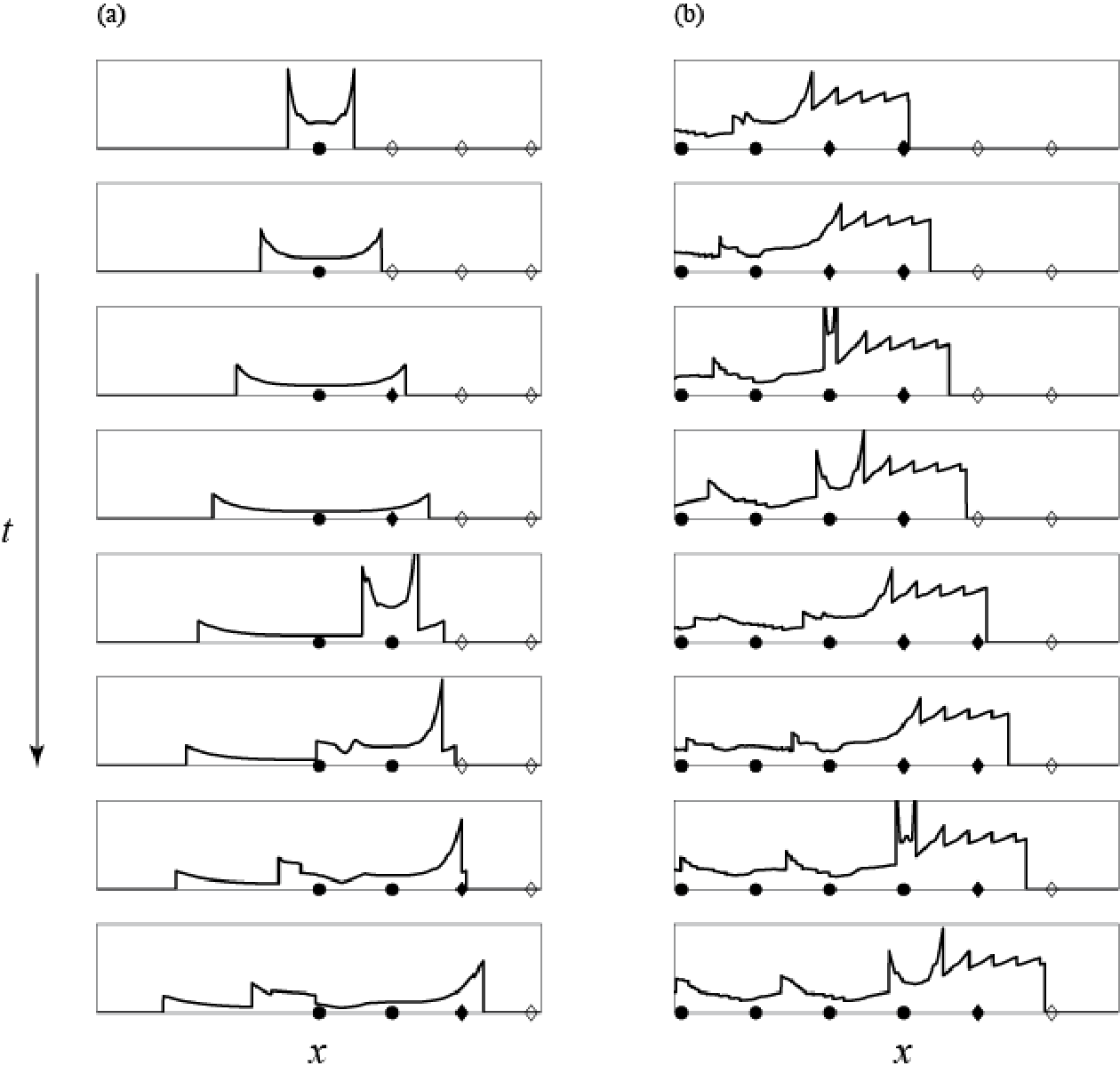}}
		\caption{Pressure profiles showing evolution of flow field with release of source energy ($\Gamma = 0.05$, $\tau = 1.5$, $Q = 50$, $\gamma = 5/3$, and lab-fixed sources)  ($a$) Triggering of the first $3$ sources and ($b$) later time evolution after the trigger of $280$ sources. The symbols plotted on the horizontal axis indicate the location of energy sources. An open diamond represents a source before being shocked, a solid diamond a source undergoing delay period after being shocked, and a solid circle a source after releasing its energy.}
	\label{Fig3}
\end{figure}

\begin{figure}
\centerline{\includegraphics[width=1.0\textwidth]{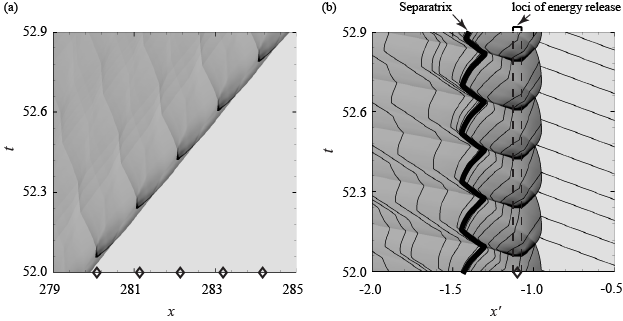}}
		\caption{$x$-$t$ diagram of flow field in ($a$) lab-fixed and ($b$) wave-fixed reference frame($\Gamma = 0.05$, $\tau = 0.1$, $Q = 50$, $\gamma = 5/3$, and lab-fixed sources). The curves in ($b$) are right-running characteristics generated by integrating the trajectory of an acoustic wave through the already-computed flow field, and the thickened curve indicates the separatrix. The symbols plotted on the horizontal axis indicate the location of energy sources.}
	\label{Fig4}
\end{figure}

The results of a sample calculation are shown in figure~\ref{Fig3}, showing the pressure profile of the computational domain for a simulation with the following parameters: $\Gamma = 0.05$, $\tau = 1.5$, $Q = 50$, and $\gamma = 5/3$. For this particular simulation, the sources were held fixed in space (i.e., they were not convected with the shock-accelerated flow).  Both early time (the first $3$ sources) and later time (after the triggering of $280$ sources) are shown in parts ($a$) and ($b$), respectively, and symbols indicate the location and status of the sources (i.e., unshocked, shocked with delay time elapsing, and energy released).  The local shock dynamics following the initial release of energy resembles a classical blast wave profile.  At later times (figure~\ref{Fig3}($b$)), a large number of saw-toothed waves, which are residual, decaying blast waves from earlier sources, can be seen superimposed upon the flow field.
Figure~\ref{Fig4} shows $x$-$t$ diagrams constructed from simulation results ($\Gamma = 0.05$, $\tau = 0.1$, $Q = 50$, $\gamma = 5/3$, and sources held fixed), with the shading indicating pressure in the flow field obtained over the region from the $280^\mathrm{th}$ to the $284^\mathrm{th}$ source. The dark V-shaped regions in the pressure fields are the blast waves generated by the energy release of the sources, propagating both forward and backward in the flow. The forward propagating blast generated by a newly triggered source catches up and accelerates the leading shock ahead of it. After the leading shock has traveled a sufficiently long distance, the process of new sources releasing their energy and the forward propagating blast waves catching up to the leading shock becomes nearly periodic. In figure~\ref{Fig4}($b$), the spatial coordinates are transformed into a reference frame moving at the average velocity of the leading shock, i.e., $x'=x-V_\mathrm{avg}t$. Right-running characteristics are constructed in figure~\ref{Fig4}($b$) by integrating an ODE describing the path of along which acoustic signals move through the domain at the particle velocity (with respect to the wave-attached frame) plus the local sound speed. The use of $x$-$t$ diagrams and characteristics analysis was first explored by \citet{McVeyToong1971} and recently by \citet{KasimovStewart2004} and \citet{Leung2010} to illustrate the physical mechanisms governing the instabilities in hypersonic exothermic flow and pulsating detonations, respectively. This construction of the $x$-$t$ diagram is a post-processing exercise performed upon the already-computed solution. In figure~\ref{Fig4}($b$), there is a characteristic (plotted as a thickened line) such that the characteristics upstream (on the right) of it can eventually reach the leading shock, while those downstream (on the left) can never reach the leading shock. This limiting characteristic is known as the separatrix, such that only the flow field upstream of it is able to influence the shock front. It can be clearly seen in figure~\ref{Fig4}($b$) that the separatix oscillates around a location which is significantly downstream from the loci of sources releasing energy.

\begin{figure}
\centerline{\includegraphics[width=1.0\textwidth]{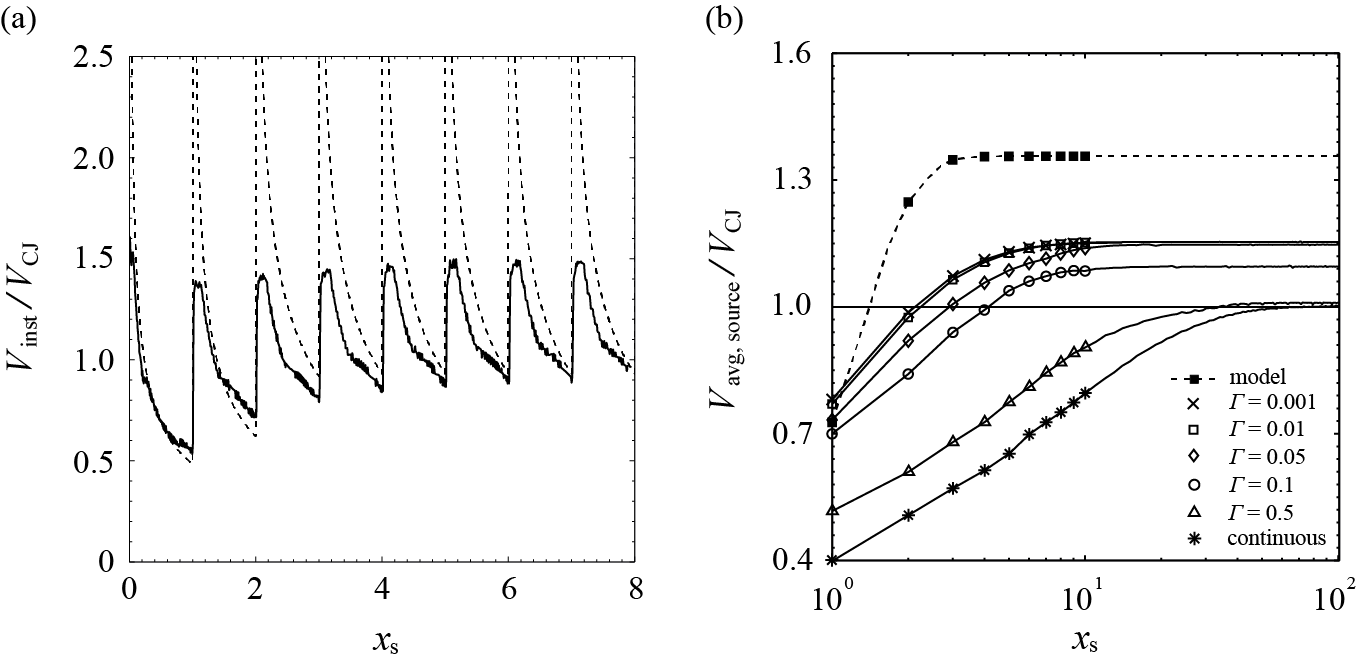}}
		\caption{The history of ($a$) the instantaneous wave velocity normalized by $V_\mathrm{CJ}$ over the first $8$ lab-fixed sources for the case of $\Gamma = 0.05$, $\tau = 0.01$, $Q = 50$, and $\gamma = 1.1$, and ($b$) the source-to-source average wave velocity over $100$ lab-fixed sources for the cases of $\Gamma = 0.001$ to $1$, $\tau = 0.01$, $Q = 50$, and $\gamma = 1.1$ as a function of the leading shock position $x_\mathrm{s}$. The heuristic model prediction is plotted as a dashed curve. In ($b$), the symbols are plotted only for the first $10$ sources.}
	\label{Fig5}
\end{figure}

Figure~\ref{Fig5}($a$) shows the evolution of the instantaneous velocity, $V_\mathrm{inst}$, normalized by $V_\mathrm{CJ}$, at which the leading wave front propagates over the first $8$ lab-fixed sources for the case of $\Gamma = 0.05$, $\tau = 0.01$, $Q = 50$, and $\gamma = 1.1$. As described in \S~\ref{Sec_Numerical_Method}, the trajectory of the leading shock front is tracked in the simulations. $V_\mathrm{inst}$ as a function of $x_\mathrm{s}$ can thus be obtained by numerically differentiating $x_\mathrm{s}(t)$ using a central difference scheme. $V_\mathrm{inst}$ exhibits large fluctuations as the successive sources are triggered and the blast waves they drive reach the leading shock front. Each time a blast wave reaches and merges with the shock front, $V_\mathrm{inst}$ is brought to a peak value, and then, decays to a local minimum velocity just before the next blast wave reaches the front. In figure~\ref{Fig5}($a$), it can be observed that both the local minimum and maximum $V_\mathrm{inst}$, just before and after the instant of the next blast wave reaching the front respectively,  monotonically increase to steady values as the sources are successively triggered. Also shown in figure~\ref{Fig5}($a$) is the prediction of a simple heuristic model that is based on the classic similarity solution of Taylor and Sedov for planar blast waves, the concept of energy partitioning of new source energy released at the blast front, and displacement of the flow field resulting from the residual influence of prior sources; see the Appendix for a complete development of this model.

With $x_\mathrm{s}(t)$ obtained from the simulation results, the average velocity at which the leading wave front travels from one source to the next can be calculated.  Figure~\ref{Fig5}($b$) shows the evolution of the source-to-source average wave velocity, $V_\mathrm{avg,source}$, as a function of the distance traveled by the leading wave front. The results shown in this figure are for the cases of various values of discreteness ($\Gamma=0.001$ to continuous), $\tau = 0.01$, $Q = 50$, and $\gamma = 1.1$ over $100$ lab-fixed sources. For all values of $\Gamma$, $V_\mathrm{avg,source}$ gradually increases and converges to a nearly constant value. A plateau value of $V_\mathrm{avg,source}$ indicates that the dynamics of the shock front triggering the energy release of the sources and the resulting blast waves reaching and supporting the shock front have reached a quasi-periodic state with a constant time interval between the trigger of subsequent sources. As the sources become more discrete ($\Gamma \to 0.001$), the plateau $V_\mathrm{avg,source}$ converges to a value that is $15\%$ in excess of the CJ velocity of the equivalent homogeneous system (note the horizontal line indicating CJ velocity) for $\gamma=1.1$. The distance required for $V_\mathrm{avg,source}$ to reach a quasi-steady value decreases from the order of $100$ for continuous energy source to that in the order of $10$ for $\Gamma = 0.001$. The heuristic model (see Appendix) prediction of how $V_\mathrm{avg,source}$ evolves as sources are triggered successively is also shown in figure~\ref{Fig5}($b$). In order to show the results for a large range in the value of source discreteness, the simulation results shown in figure~\ref{Fig5}($b$) are those obtained by using the adaptive mesh refinement code.

The following subsections systematically study the average velocity of the wave as the various parameters in the model are varied.  The influence of the source scenarios (lab-fixed vs. convected sources) is also investigated.  All average velocities, subsequently reported in this paper and denoted as $V_\mathrm{avg}$, were measured after the leading shock front has reached a quasi-steady value. The length of the simulation domain, i.e., the total number of discrete sources, over which the detonation wave needs to propagate to reach a quasi-periodic state, varied approximately from 20 to 100 as $\Gamma$ increases from $0.001$ to $1$. In the simulations using the fixed, uniform grid code, which are the cases with relatively less discrete sources ($\Gamma \geq 0.05$), the leading front propagated through $300$ sources, and the average velocities calculated over the last $100$, $50$, or $10$ sources differed by less than $0.1\%$. Hence, for the simulations using the fixed, uniform grid code, average velocity over the last $50$ sources, i.e., from the $250^\mathrm{th}$ to $300^\mathrm{th}$ source, is reported in the following subsections of \S~\ref{Sec_Results} as $V_\mathrm{avg}$.  For the cases with extremely discrete sources ($\Gamma \leq 0.01$) which were only simulated using the adaptive mesh code, the length of the simulation domain cannot be extended beyond $30$ sources due to computational time and memory limits. This length is, however, sufficient for the wave propagation to reach a quasi-periodic state as shown in figure~\ref{Fig5}($b$).  Hence, for those cases, it is sufficient to report the average velocity over the last $10$ sources as $V_\mathrm{avg}$. All values are normalized by the CJ velocity of the equivalent homogenized media.

\subsection{Fixed Sources}
\label{Sec_Fixed_Sources}

\subsubsection{Discreteness}
\label{Sec_Discreteness}
The effect of the spatial concentration of the sources, i.e., the discreteness parameter $\Gamma = \widetilde{W}/\widetilde{L}$, on the average shock front velocity (normalized by CJ velocity) is reported in figure~\ref{Fig6}($a$).  For these calculations, the value of $Q$ was held constant at $50$, the delay time was a constant $\tau = 0.01$, and the sources were assumed to remain fixed after the passage of the shock.  Note that in all simulations, as the width of the source was expanded to fill the entire space between sources ($\Gamma \to 1$), the average wave velocity converged to the CJ velocity (to within $0.5\%$). Also note that in this limit, the energy of each source is still released instantaneously, only now there is no gap between subsequent sources.  The fact that the CJ velocity is recovered in the limit of homogeneous energy released in a piecewise continuous manner is good confirmation in the overall implementation of this model.

As the sources are made progressively more discrete by concentrating them in space, an increasing deviation in the average velocity away from the CJ solution is observed.  For $\gamma = 5/3$, the deviation from CJ reaches a plateau value of $V_\mathrm{avg}/V_\mathrm{CJ} = 1.025$ as $\Gamma$ decreases below $0.2$. For lesser values of $\gamma$, greater deviations from CJ are observed and the deviation continues to increase as the sources are made increasingly discrete ($\Gamma \to 0$).  For the cases with highly discrete sources, i.e., $\Gamma < 0.05$, the numerical code with a fixed grid cannot adequately resolve the sources with sufficient resolution ($100$ computational cells per source) due to computer memory limitations, motivating the use of the adaptive refinement code.  The uniform, fixed grid and adaptive mesh refinement codes exhibit good agreement, as seen by comparing the solid and open symbols in figure~\ref{Fig6}($a$), respectively, providing confidence that the results of this study, and the observed super-CJ velocities in particular, are not an artifact of the particular solver used.

For the case of $\gamma = 1.1$, the average wave speed achieves a value of $15\%$ greater than CJ (i.e., $V_\mathrm{avg}/ V_\mathrm{CJ} = 1.15$) as the sources become concentrated into $5\%$ of the available volume (i.e., discreteness $\Gamma = 0.05$).  This is the greatest deviation away from the CJ solution observed in this study.  The effect of the ratio of specific heats is explored further in the next section.  As the sources are made progressively more discrete ($\Gamma < 0.05$) in the $\gamma = 1.1$ case, the average wave speed appears to reached a plateau value.  We interpret this plateau as an asymptotic limit in spatial discreteness, and we explore the concept of an asymptotic limit of $\delta$-function-like sources further in Appendix~\ref{appA}.

\subsubsection{Specific Heat Ratio}
\label{Sec_Specific_Heat_Ratio}
\begin{figure}
\centerline{\includegraphics[width=1.0\textwidth]{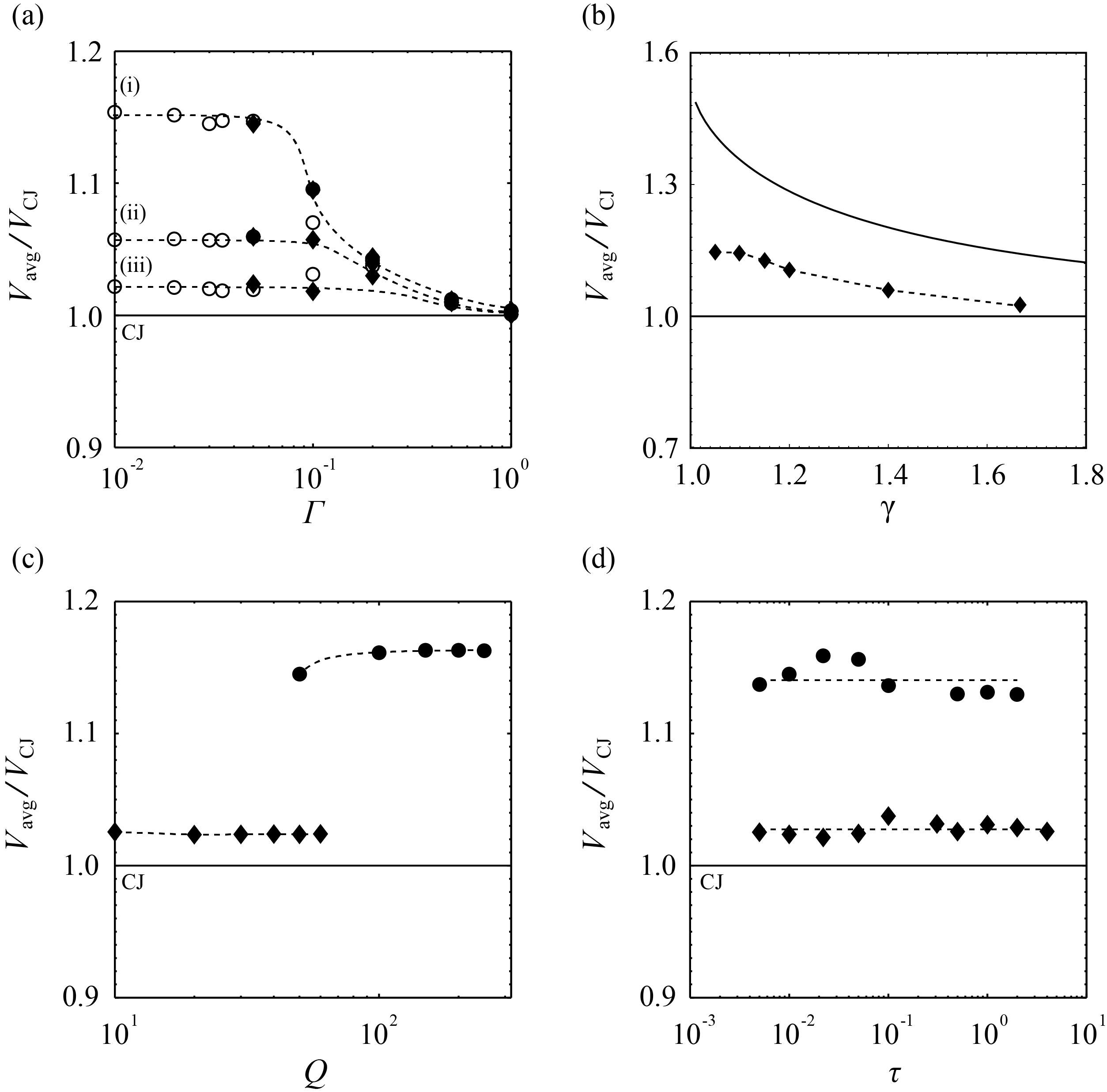}}
		\caption{Simulation results of the average wave velocity normalized by the CJ velocity as a function of ($a$) discreteness, ($b$) specific heat ratio, ($c$) heat release, and ($d$) delay time with lab-fixed sources. Simulation data obtained using the adaptive refinement code are plotted as open circles. In ($a$), the simulation results with $\tau = 0.01$, $Q = 50$, and $\gamma = 1.1$, $1.4$ , and $5/3$ are labeled (i), (ii), and (iii), respectively. In ($b$), the simulation results with $\Gamma = 0.05$, $\tau = 0.01$, and $Q = 50$ are plotted as diamonds and compared to the heuristic model prediction plotted as a solid curve (see Appendix~\ref{appA}). In ($c$), the simulation results with $\Gamma = 0.05$ and $\tau = 0.01$ are plotted as circles and diamonds for $\gamma = 1.1$ and $5/3$, respectively. In ($d$), the simulation results with $\Gamma = 0.05$ and $Q = 50$ are plotted as circles and diamonds for $\gamma = 1.1$ and $5/3$, respectively.}
	\label{Fig6}
\end{figure}

The average wave speed measured for quasi-periodic propagation is reported as a function of the specific heat ratio $\gamma$ in figure~\ref{Fig6}($b$). For these simulations, the value of heat release was fixed at $Q = 50$, the delay time at $\tau = 0.01$, and the discreteness was fixed at $\Gamma = 0.05$. From figure~\ref{Fig6}($a$), this value of the discreteness parameter is seen to be sufficiently small that the wave is propagating in the asymptotic limit of highly discretized sources, such that decreasing the value of $\Gamma$ further will no longer influence the result.  The average velocity (as normalized by the CJ velocity) is seen to increase as the value of $\gamma$ is decreased. Also shown in this figure is the result of the heuristic model developed in Appendix~\ref{appA}. The qualitative agreement between the model and the simulations will provide a basis to explain the dependence on $\gamma$, as discussed in \S~\ref{Sec_Discussion}. 

\subsubsection{Heat Release}
\label{Sec_Heat_Release}
The value of the heat release $Q$ was varied in a way such that the Mach number of the CJ detonation in the homogenous media remained between values of $M_\mathrm{CJ} = 4$ and $M_\mathrm{CJ} = 10$, in order to be representative of real detonable mixtures.  This resulted in the value of $Q$ varying between $50$ and $250$ for $\gamma = 1.1$ and between $10$ and $60$ for $\gamma = 5/3$. The average wave speed measurements of these simulations are shown in figure~\ref{Fig6}($c$).  The observed wave speed normalized by the CJ speed is not particularly sensitive to the heat release of the mixture.

\subsubsection{Delay Time}
\label{Sec_Delay_Time}
The assumption of a fixed delay time, which is not influenced by the strength of the shock wave that triggers the source or the post-shock thermodynamic states, may be viewed as an unrealistic assumption in the model.  Thus, it is important to examine the influence of varying the delay time $\tau$ in separate calculations.  The effect of varying the delay time, which remained fixed for each simulation, is reported in figure~\ref{Fig6}($d$) for discreteness  $\Gamma = 0.05$ and $Q = 50$.  Note that, as the delay time is varied from $\tau = 0.004$ to $\tau = 3$, the observed average wave speed does not exhibit any overall trend. A small deviation is observed around $\tau = 0.03$ and $\tau = 0.1$ for $\gamma = 1.1$ and $5/3$, respectively, however, we are unable to attribute any significance to this particular result.

\subsection{Convected Sources}
\label{Sec_Convected_Sources}
\begin{figure}
\centerline{\includegraphics[width=0.6\textwidth]{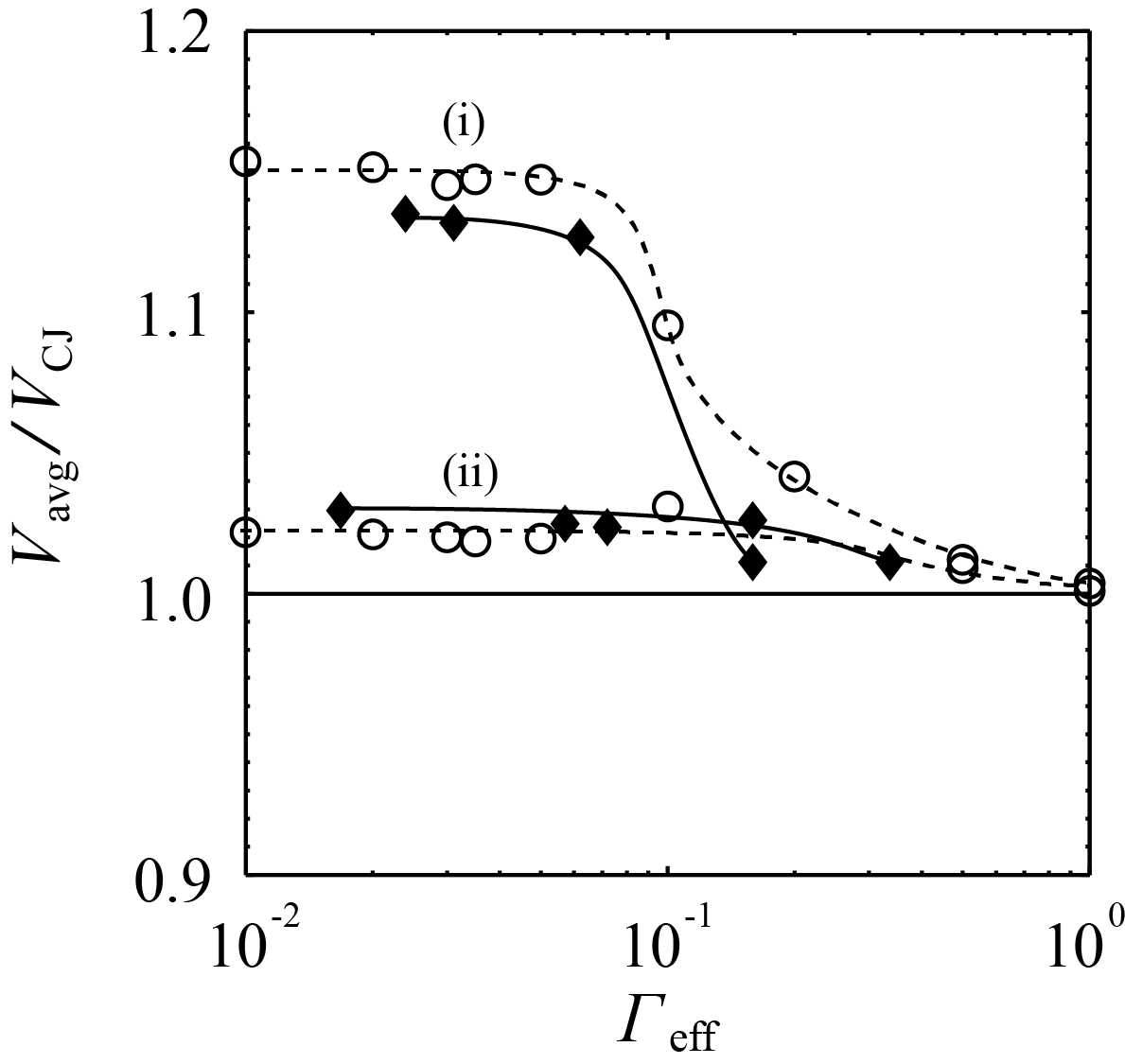}}
		\caption{Simulation results of the average wave velocity normalized by the CJ velocity as a function of effective discreteness with $\tau = 0.01$, $Q = 50$, and (i) $\gamma = 1.1$ and (ii) $5/3$. The simulation data for the cases of lab-fixed and convected sources are plotted as open circles and solid diamonds, respectively.}
	\label{Fig7}
\end{figure}

The simulations presented so far are for the case of sources that remain spatially fixed after the passage of the shock. In order to examine the influence of this assumption, a second series of simulations were performed in which the energy source is embedded in the medium through which the shock wave propagates, and following the passage of the shock wave, is convected along with the flow. Since a convected source is compressed by the shock wave, the width of a source when it releases energy after a finite delay time $t_\mathrm{d}$ is smaller than its initial width before the passage of the shock. Hence, the effective discreteness ($\Gamma_\mathrm{eff}$) is smaller than that in the initial distribution of sources. Simulations with $Q=50$, $\tau=0.01$, and $\gamma=1.1$ and $5/3$ were performed for both cases of fixed and convected sources. The measured average wave speeds of these simulations are shown in figure~\ref{Fig7}. The trend of an increase in average wave speed for the cases of convected sources, as the value of $\Gamma_\mathrm{eff} \to 0$, is observed to be similar to that for the cases of fixed sources. For $\Gamma_\mathrm{eff}$ approximately below $0.1$, the average wave speed reaches a plateau value of $15\%$ and $3\%$ greater than the CJ value for $\gamma=1.1$ and $5/3$, respectively.

\section{Analysis}
\label{Sec_Analysis}
The fact that a spectrum of quasi-periodic solutions that propagate at average speeds significantly greater than the CJ speed (i.e., super-CJ solutions) when the source energy is sufficiently concentrated in space is a result that could be met with a large degree of skepticism, given the considerable success of the CJ solution in comparison with experiments.  In order to provide an explanation of this finding, the results of select simulations reported in the prior section were analyzed using the conventional pressure/specific volume ($p$-$v$) representation of thermodynamic states.  Following the approach developed by \citet{LeeRadulescu2005}, \citet{Radulescu2007JFM}, and \citet{Sow2014JFM}, the flow field of the wave propagating in a discrete source detonation will be analyzed in a moving reference frame and the flow at each point in the domain will be averaged over time.  This will enable the average structure of the wave to be compared to the classical structure of the ZND model of a detonation, as visualized in the $p$-$v$ diagram.

Similar to the analysis performed by \citet{Radulescu2007JFM}, the averaging is done in a reference frame moving at the average velocity of the wave calculated from the $250^\mathrm{th}$ to the $300^\mathrm{th}$ source. Thus, in the moving reference frame, the spatial coordinate and particle velocity are transformed as $x'=x-V_\mathrm{avg}t$ and $u'=u-V_\mathrm{avg}$, respectively. For convenience, $u$ denotes the particle velocity with respect to the moving frame in this section.

A simple time averaging, or Reynolds averaging procedure, is then applied to density and pressure as follows
\begin{equation}
\bar{\rho}\left( x' \right) = \frac{1}{t_{2}-t_{1}} \int_{t_{1}}^{t_{2}} \rho \left( x',t \right) \mathrm{d}t
\label{Eq8}
\end{equation}
\begin{equation}
\bar{p}\left( x' \right) = \frac{1}{t_{2}-t_{1}} \int_{t_{1}}^{t_{2}} p \left( x',t \right) \mathrm{d}t
\label{Eq9}
\end{equation}
where $t_1$ and $t_2$ indicate the starting and ending time of the period, respectively, over which $\rho$ and $P$ are averaged. The bar ``$\overline{~~}$'' indicates time-averaged variables. Favre averaging (i.e., density weighted averaging) is applied to particle velocity and specific energy as follows,
\begin{equation}
u^\ast = \frac{\overline{\rho u}}{\bar{\rho}} \;\; \mathrm{and} \;\; u = u^\ast+u''
\label{Eq10}
\end{equation}
\begin{equation}
e^\ast = \frac{\overline{\rho e}}{\bar{\rho}} \;\; \mathrm{and} \;\; e = e^\ast+e''
\label{Eq11}
\end{equation}
where superscripts ``$\ast$'' and $''$ indicate Favre-averaged variables and fluctuating quantities, respectively. The average structure of the wave is therefore governed by the one-dimensional, stationary Favre-averaged Euler equations as follows, 
\begin{equation}
\frac{\partial}{\partial x'} \left( \bar{\rho} u^\ast \right) = 0
\label{Eq12}
\end{equation}
\begin{equation}
\frac{\partial}{\partial x'} \left( \bar{\rho} {u^\ast}^2 + \bar{p} + \overline{\rho {u''}^2}\right) = 0
\label{Eq13}
\end{equation}
\begin{equation}
\frac{\partial}{\partial x'} \left( \bar{\rho} e^\ast u^\ast + \bar{\rho}\left(e'' u'' \right)^\ast + \overline{pu} \right) = 0
\label{Eq14}
\end{equation}
For an averaged fluid element traversing the wave structure, its averaged mass and momentum are conserved according to
\begin{equation}
\overline{M}^2 = \frac{\bar{p}-1+f}{\gamma\left( \bar{v}-1 \right)}
\label{Eq15}
\end{equation}
where $\bar{v} = 1/\bar{\rho}$ is the averaged specific volume, $\overline{M}=V_\mathrm{avg}/c_0$ the Mach number of the average wave velocity, and $f=\overline{\rho {u''}^2}$ is the Reynolds stress, which measures the intensity of fluctuations in momentum. Note that when $f=0$, i.e., there is no fluctuation in fluid momentum over time, (\ref{Eq15}) reverts to the ideal Rayleigh line that a fluid element traversing a one-dimensional steady detonation structure is expected to follow.

The average sound speed, which is assumed to be independent of the intensity of fluctuation, can be calculated as
\begin{equation}
c^\ast = \sqrt{\frac{\gamma \bar{p}}{\bar{\rho}}}
\label{Eq16}
\end{equation}
The sonic point in the one-dimensional averaged wave structure is located at the position at where $u^\ast + c^\ast = 0$. The importance of using a mean steady detonation profile to determine the location of the effective sonic plane was highlighted by \citet{LeeRadulescu2005}.

\begin{figure}
\centerline{\includegraphics[width=1.0\textwidth]{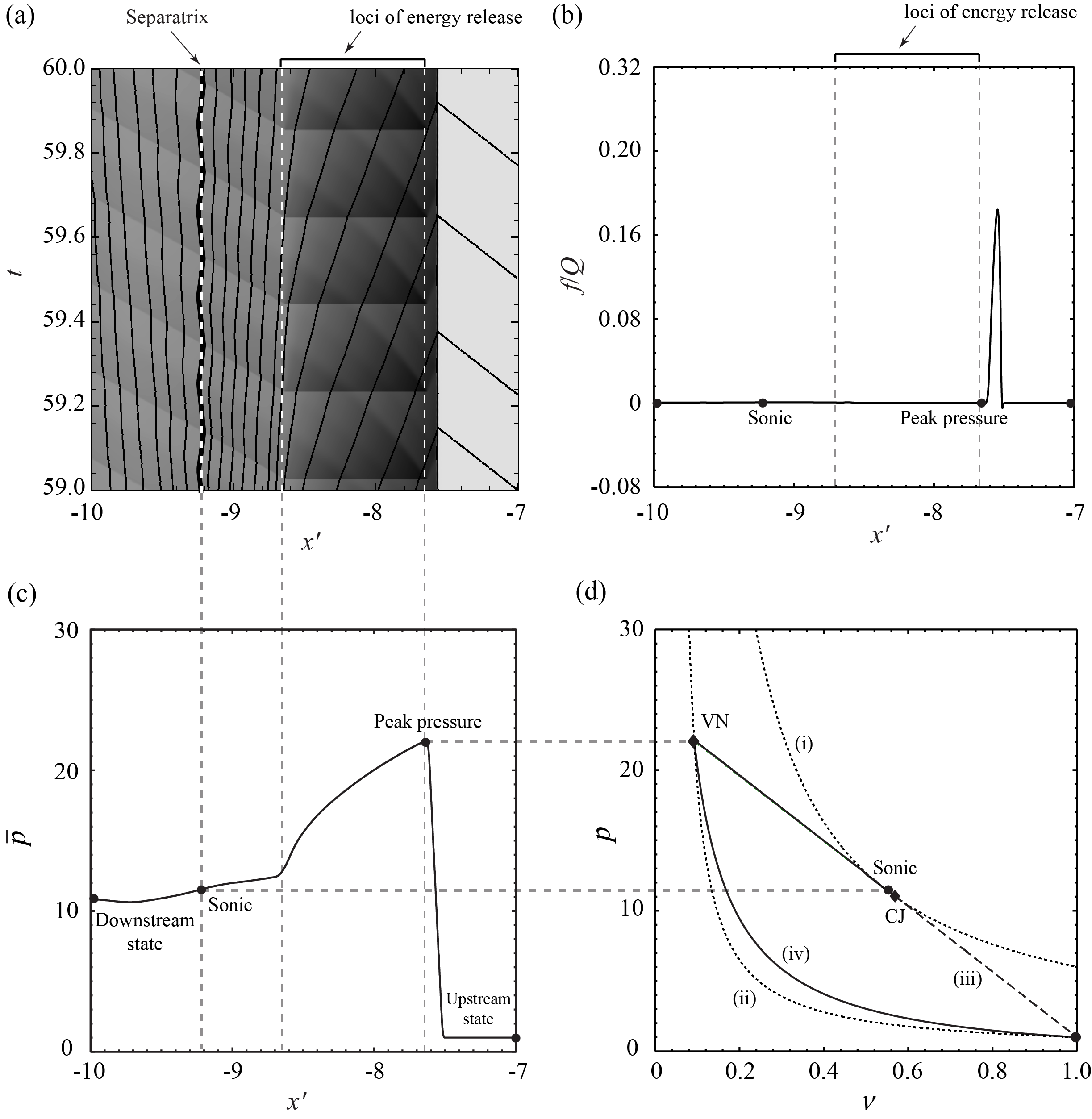}}
		\caption{For the case of continuous (i.e., $\Gamma=1$) lab-fixed energy sources, $Q=50$, $\tau = 0.1$, and $\gamma = 5/3$, ($a$) pressure contour plotted as a $x'$-$t$ diagram showing the evolution of the flow field and right-running characteristics, ($b$) the spatial profile of the intensity of mechanical fluctuation $f$ normalized by $Q$, ($c$) the one-dimensional steady structure of the time-averaged pressure $\bar{p}$ in the wave-attached reference frame, and ($d$) the $p$-$v$ representation of the one-dimensional averaged wave structure. In ($d$), curve (i) is the equilibrium Hugoniot curve of complete energy release $Q=50$, curve (ii) is the Hugoniot curve of zero energy release $Q=0$, curve (iii) is the ideal Rayleigh line corresponding to a leading shock strength of $M_\mathrm{CJ}$, and curve (iv) is the $p$-$v$ thermodynamic path taken by a fluid element traversing the one-dimensional averaged wave structure.}
	\label{Fig8}
\end{figure}

Figure~\ref{Fig8} shows the simulation results for the case of continuous (i.e., $\Gamma =1$) lab-fixed energy sources, $Q=50$, $\tau = 0.1$, and $\gamma = 5/3$ analyzed using the Favre averaging approach and $p$-$v$ representation of thermodynamic states. Note that while the source energy is uniform throughout the medium, it is still released in piecewise-continuous segments with width $L$, which is non-dimensionalized to a value of unity when reported here. In figure~\ref{Fig8}($a$), the evolution of the flow field is plotted as a grayscale contour of pressure in a $x'$-$t$ diagram. Figure~\ref{Fig8}($b$) shows the spatial profile of $f$ in the wave-attached reference frame, which measures the intensity of mechanical fluctuations around the Favre-averaged value in the flow field, non-dimensionalized by $Q$. The one-dimensional steady structure of the time-averaged pressure $\bar{p}$ in the wave-attached reference frame is plotted in figure~\ref{Fig8}($c$) below the $x'$-$t$ diagram. The thermodynamic path taken by a fluid element traversing this one-dimensional steady, averaged wave structure is plotted in a $p$-$v$ diagram as shown in figure~\ref{Fig8}($d$). The location of the sharp increase in $\bar{p}$ from the upstream initial state to its peak value occurs in the region between the leading shock and the loci of the source releasing energy in the $x'$-$t$ diagram. The peak in the averaged pressure matches the von Neumann pressure, i.e., the downstream pressure behind a normal shock of strength $M_\mathrm{CJ}$, shown in figure~\ref{Fig8}($d$). As shown in figure~\ref{Fig8}($a$), the $u+c$ characteristics near the separatrix are quasi-straight lines with slight wiggles as they pass through backward-running shock waves and contact surfaces.

Comparing figure~\ref{Fig8}($a$) and ($c$), the sonic point in the averaged profile is found to match the trajectory of the limiting characteristic of the separatrix, which appears downstream from the loci of sources releasing energy. The thermodynamic state at this averaged sonic point nearly coincides with the CJ state of a one-dimensional steady detonation wave propagating at $M_\mathrm{CJ}$, as seen in figure~\ref{Fig8}($d$). This is consistent with the result that the $\overline{M}$ obtained from the simulation with continuous energy sources is only $0.3\%$ greater than the $M_\mathrm{CJ}$ for the same total energy release $Q$. Since the CJ point is the intersection of the ideal Rayleigh line and the equilibrium Hugoniot curve of complete energy release, the averaged flow field relaxes to equilibrium at the sonic point.\footnote{It is important to note that, in this problem, the term ``equilibrium'' refers to mechanical (or hydrodynamic) equilibrium that the flow reaches at the sonic point in the classical CJ solution, rather than chemical equilibrium of a system with reversible reactions.} In figure~\ref{Fig8}($b$), it can be seen that mechanical fluctuation $f/Q$ is significantly positive only in a narrow region which corresponds to the sharp increase in averaged pressure shown in figure~\ref{Fig8}($c$). In figure~\ref{Fig8}($d$), the averaged $p$-$v$ path from the upstream initial state to the peak averaged pressure (or the von Neumann) state significantly departs from the ideal Rayleigh line corresponding to a wave speed of $M_\mathrm{CJ}$, which can be interpreted as the fact that the mechanical fluctuation is concentrated between the shock and energy release as shown in figure~\ref{Fig8}($b$). As $f/Q$ rapidly decays to and remains at zero downstream of the location of peak averaged pressure to the sonic point, the averaged $p$-$v$ path connecting the peak averaged pressure state to the sonic point coincides with the ideal Rayleigh line. In the case of continuous energy sources, despite the mechanical fluctuation caused by the interaction between the forward-running blast generated by each individual source releasing its energy and the preceding shock front, the highly unsteady wave propagation and energy deposition process has a temporal average structure, which is nearly equivalent to a steady one-dimensional detonation structure with the same amount of heat release.  The excellent agreement obtained between the averaged structure of the wave extracted from the simulations (with continuous energy release) and the steady ZND structure can be taken as validation of the averaging method used in this study.

\begin{figure}
\centerline{\includegraphics[width=1.0\textwidth]{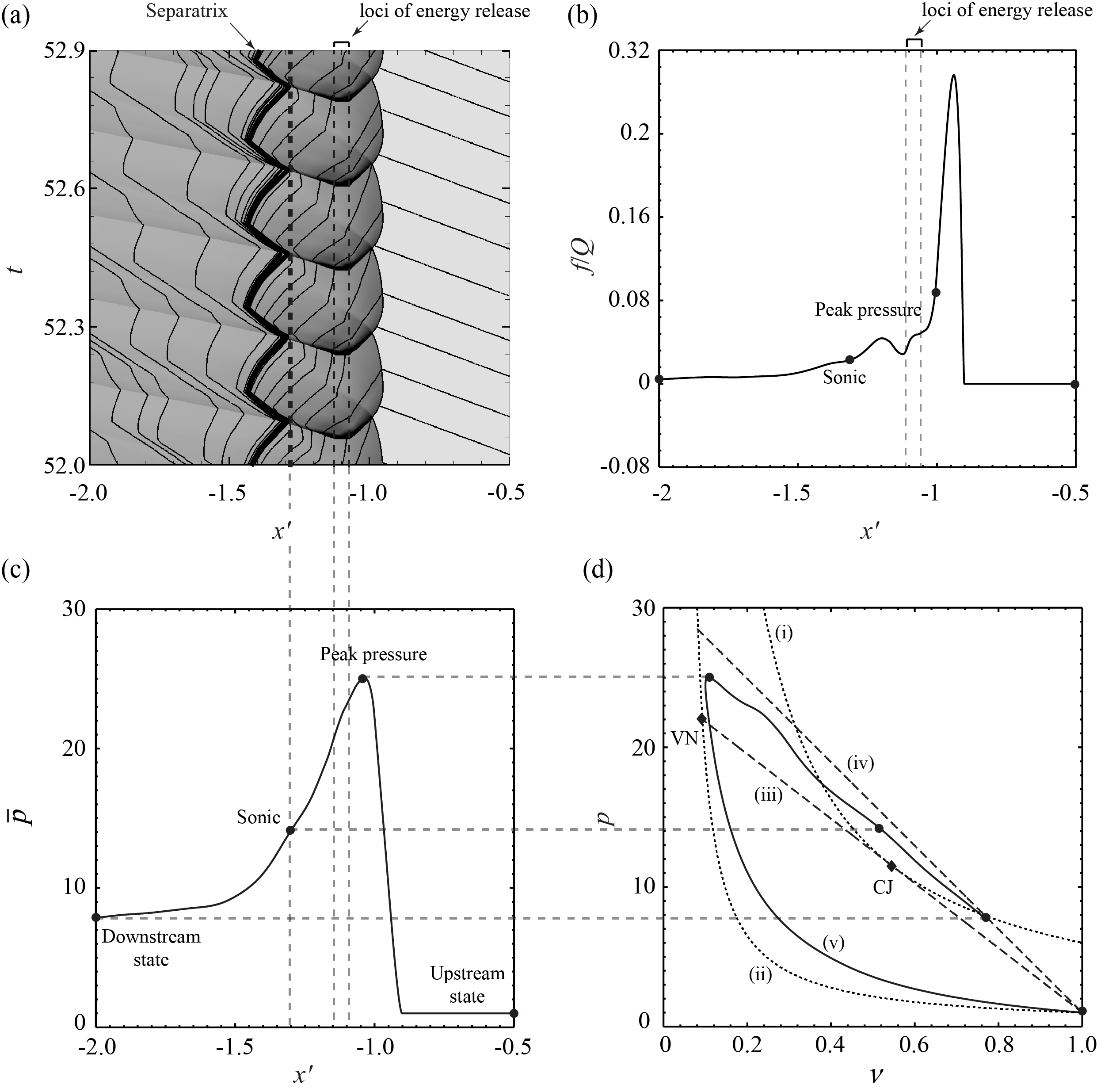}}
		\caption{For the case of highly discrete ($\Gamma =0.05$) lab-fixed energy sources, $Q=50$, $\tau = 0.1$, and $\gamma = 5/3$, ($a$) pressure contour plotted as a $x'$-$t$ diagram showing the evolution of the flow field and right-running characteristics, ($b$) the spatial profile of the intensity of mechanical fluctuation $f$ normalized by $Q$, ($c$) the one-dimensional steady structure of the time-averaged pressure $\bar{p}$ in the wave-attached reference frame, and ($d$) the $p$-$v$ representation of the one-dimensional averaged wave structure. In ($d$), curve (i) is the equilibrium Hugoniot curve of complete energy release $Q=50$, curve (ii) is the Hugoniot curve of zero energy release $Q=0$, curves (iii) and (iv) are the ideal Rayleigh lines corresponding to a leading shock strength of $M_\mathrm{CJ}$ and $\overline{M}$, respectively, and curve (v) is the $p$-$v$ thermodynamic path taken by a fluid element traversing the one-dimensional averaged wave structure.}
	\label{Fig9}
\end{figure}

Figure~\ref{Fig9} shows the simulation results for the case of highly discrete ($\Gamma =0.05$) lab-fixed energy sources, $Q=50$, $\tau = 0.1$, and $\gamma = 5/3$ analyzed using the Favre averaging approach and the $p$-$v$ representation of thermodynamic states. As seen from figure~\ref{Fig6}($b$), $\Gamma=0.05$ approximately corresponds to the limit of discreteness where the wave velocity becomes independent of the value of $\Gamma$ (i.e., making the sources even more discrete does not result in greater deviation from the CJ solution). The $x'$-$t$ diagram of flow field evolution, the spatial profile of $f/Q$, the one-dimensional steady profile of $\bar{p}$, and the $p$-$v$ diagram of the averaged wave structure are plotted in figure~\ref{Fig9}($a$), ($b$), ($c$), and ($d$), respectively. Comparing figure~\ref{Fig9}($a$) and ($c$), it can be found that the increase in averaged pressure from the initial state to its peak value spans over the region where the forward-running blast waves interact with the leading shock front in the $x'$-$t$ diagram. As shown in figure~\ref{Fig9}($b$), the maximum intensity of mechanical fluctuation is associated with wave interactions in this region. 

In contrast to the continuous case (figure~\ref{Fig8}), as shown in figure~\ref{Fig9}($a$), the $u+c$ characteristics exhibit a ``zig-zag'' pattern as they are processed by much stronger backward-running blast waves. The trajectory of the limiting characteristic (i.e., separatrix) oscillates back and forth while its overall zig-zag pattern remains in a fixed region relative to the leading wave front over time. The sonic point found in the one-dimensional Favre-averaged wave structure coincides with the right (or upstream) end of the limiting characteristic. Similar to the continuous case, the location of the sonic point is significantly downstream from the loci of discrete sources releasing energy. Plotting the $p$-$v$ path taken by a fluid element traversing the Favre-averaged wave structure in figure~\ref{Fig9}($d$), it can be seen that the peak averaged pressure is greater than the von Neumann pressure behind a shock of strength $M_\mathrm{CJ}$. The averaged sonic point departs from the CJ point, and is not found on the equilibrium Hugoniot curve for $Q=50$. Hence, equilibrium is not reached as the separatrix and the effective sonic surface in the flow field are encountered.

Equilibrium is eventually attained in the far field downstream from the sonic point, as this downstream state coincides with the intersection between the equilibrium Hugoniot curve of $Q=50$ and the Rayleigh line of a wave speed $\overline{M}$. The averaged pressure at this equilibrium state is significantly less than that of the CJ state. The $\overline{M}$ for $\Gamma =0.05$ discrete case, which is $14\%$ greater than $M_\mathrm{CJ}$, is associated with a weak detonation solution. As shown in figure~\ref{Fig9}($b$), the mechanical fluctuation retains a significantly positive value from the leading wave front to a location beyond the sonic point, such that the entire averaged $p$-$v$ path departs from the ideal Rayleigh line based on a one-dimensional steady detonation structure. The quantity $f/Q$ eventually decays to zero in the far field downstream, consistent with the result that the far-field downstream thermodynamic state reaches the intersection between the equilibrium product Hugoniot curve and the Rayleigh line associated with the wave speed $\overline{M}$.

\section{Discussion}
\label{Sec_Discussion}
The results presented in this paper demonstrate that the classical CJ solution for detonation velocity breaks down and is no longer valid as the source of energy is spatially concentrated into highly discretized sources.  In order to explain this result, the analysis in \S~\ref{Sec_Analysis} shows that flow crossing the separatrix (i.e., the limiting characteristic that can influence the shock front) has not reach equilibrium due to on-going shock wave interactions that are occurring in the flow for the case of a highly discretized medium.  This separatrix is the unsteady flow analog to a sonic surface, and the fact that the flow is not in equilibrium at the effective sonic surface is a hypothesis as to why non-CJ speeds are observed.  As seen in figure~\ref{Fig9}($d$), the flow does eventually reach the equilibrium product Hugoniot downstream of the effective sonic surface.  This result suggests that the super-CJ speeds in this study are properly designated as weak detonations.  Prior examples of weak detonations include so-called pathological detonations, in which competing exothermic and endothermic reactions permit the flow in the reaction zone to pass through a sonic point while in a non-equilibrium state \citep{von1942,Fickett2000,HigginsChapter2}, and ultrafast detonations \citep{AndersonLong2003}, in which chemical reactions are sufficiently fast that significant reaction occurs within the shock front itself, enabling the wave to by-pass the von Neumann state and proceed directly to the weak branch of the product Hugoniot. The mechanism suggested in this paper, namely, a high degree of flow non-equilibrium resulting from shock interactions, is believed to be a new mechanism able to realize weak (non-CJ) solutions.

The deviations from CJ speed depend significantly upon the ratio of specific heats, $\gamma$, as seen in figure~\ref{Fig6}($b$).  The heuristic model developed in the Appendix, which examines the limiting case of the time delay going to zero ($\tau \to 0$) and the spatial discreteness also approaching zero ($\Gamma \to 0$), such that the energy release of the first source will result in a classical point blast solution, may provide an explanation for this trend. In this model, subsequent sources will release their energy on the shock front of the prior source, and as discussed in the Appendix, the partitioning of energy release of that source into forward and backward directed blast waves is related to the density ratio in the limit of a strong shock front ($\frac{\gamma + 1}{\gamma - 1}$).  Since this function increases rapidly as the value of gamma approaches unity ($\gamma \to 1$), we would expect the energy release of subsequent sources to be increasingly directed in the direction of propagation.  In simpler language, in the limit of $\gamma$ going to unity, the strong shock front becomes a solid wall, so that energy released on that wall results in a blast wave directed entirely forward.  This picture may provide a qualitative explanation for the influence of $\gamma$, and indeed the simple heuristic model developed in the Appendix performs better in predicting the average wave speed than might be expected, as seen in figure~\ref{Fig6}($b$).

As the value of dimensionless heat release $Q$ varies in a way such that the CJ detonation velocity for an equivalent homogeneous reactive medium remains in the range between $M_\mathrm{CJ} = 4$ and $M_\mathrm{CJ} = 10$, the ratio of the resulting $V_\mathrm{avg}$ in the discrete reactive system to the homogenous equivalent $V_\mathrm{CJ}$ exhibits no significant change (see figure~\ref{Fig6}($c$)). An insight obtained from the heuristic model is that the key dynamics of the wave propagation in a medium of discrete sources can be fairly well pictured as a complex superposition of blast waves generated by a series of point sources. According to the Taylor-Sedov blast wave solution, the average speed at which the blast wave front propagates over a fixed distance has a square root dependence on the energy released by a planar source, namely, $\sqrt{Q}$. In the limit of large heat release, the CJ velocity also linearly depends on $\sqrt{Q}$. Hence, this dependence of $\sqrt{Q}$ might be expected to cancel out when $V_\mathrm{avg}$ is normalized by $V_\mathrm{CJ}$, and this may explain the results presented here.

The result that the deviations of the average wave speed away from the CJ speed do not depend on delay time is another result from this study, as seen in figure~\ref{Fig6}($d$). The delay time was varied between $\tau=0.004$, meaning that the sources release their energy effectively instantaneously upon contact with the leading shock, and $\tau=3$, meaning that the blast wave release by a source must approximately propagate a distance of three source spacings before it reaches the location where the shock was when the source released its energy, by which time the leading shock had moved further downstream as well, resulting in it taking even longer for the blast to reach the leading shock.  In this limit of large $\tau$, given the large number of blast wave interactions that occur before the blast from a source actually reaches the leading shock and contributes to its sustenance, it is perhaps remarkable that the average speed does not depend upon the delay time. Note, however, that in the classical CJ detonation model, the details of the reaction zone, such as kinetic rates, do not affect the detonation wave velocity. In this connection, it is perhaps plausible that the results do not depend on the delay time of the discrete sources.

The results of this study may have application to resolving the conundrum that was discussed in the Introduction (\S~\ref{Sec_Intro}), namely, resolving why the CJ criterion is so successful in describing the average propagation velocity of transient and multidimensional detonation waves. Unstable detonations feature pockets of gas that may get compressed by multiple shock waves and explode, while other portions of the detonation cell might be described by a non-reacting shock wave. The study of \citet{KiyandaHiggins2013} estimated that nearly the entire second half of a detonation cell in low pressure methane/oxygen does not react coupled to the shock front, but rather burns as shock-compressed pockets detached from the leading front.  The effective discreteness of such unstable detonations, as defined in this paper, is unlikely to be less than $\Gamma = 0.3$. As shown in figure.~\ref{Fig7}, it is unlikely that deviations from the ideal CJ speed would be more than a few percent for this degree of discreteness, a difference that is hardly experimentally resolvable.  This result may also have relevance to the phenomenon of galloping detonation, wherein steady detonation propagation is not possible due to the tube diameter being smaller than a characteristic cell size required for propagation, such that the detonation fails to approximately half CJ speed and periodically reinitiates to an initially overdriven wave via a process similar to deflagration to detonation transition (DDT).  Despite the fact that the cycle of the galloping detonation occurs over hundreds of tube diameters and the wave should experience significant heat and momentum losses due to the relatively small size of the tube used, galloping detonations continue to propagate remarkably close to the CJ speed on average \citep{LeeDupre1995,GaoNgLee2015,Jackson2015ICDERS}.  This aspect of the discrete source detonation model has been further explored by \citet{Radulescu2015APS}. The possibility that all real detonations might be weak detonations (if only slightly away from the CJ solution) is intriguing, as this idea was suggested by \citet{Davis1987} but has not attracted significant attention in detonation research.

The results of the present study are fundamentally different from those of \citet{Mi2015}, which examined the discrete source detonation problem in a Burgers analog system. The one-dimensional, inviscid scalar Burgers equation with a source term was proposed by \citet{Fickett1979} and \citet{Majda1981} as an analog to the reactive Euler equations in order to explore detonation dynamics. In recent years, study of this scalar analog system has generated a number of interesting results, including the existence of pulsating, chaotic solutions \citep{Radulescu2011PRL,Kasimov2013PRL,Tang2013CS,Faria2015CS}. The study of \citet{Mi2015} was a precursor to the present study using the inviscid Burgers equation with periodically spaced $\delta$-function sources that were triggered by the passage of the leading shock front. Both regularly spaced sources with fixed delay and randomly spaced sources with randomly generated delays were considered. The resulting wave dynamics involved the interaction of a number of sawtooth-profiled blast waves generated by the sources, which for the Burgers equation the trajectories of the blast waves could be solved analytically. In all cases considered, the average propagation velocity was found to be within $1\%$ of the CJ detonation velocity of the equivalent CJ system. That result is in seeming contradiction to the results of the present study, which exhibited super-CJ solutions, so comparing these two studies warrants further discussion. The Euler equations have three families of propagating characteristics, namely, right-running, left-running, and particle characteristics, while the Burgers equation has only the family of right-running characteristics. This feature of the Burgers equation system might explain the unusual dynamics noted in \citet{Mi2015} (see \S~\ref{Sec_Convected_Sources} and figure~\ref{Fig7}($b$)) wherein the limiting characteristic became coincident with the locus of new sources in the case of regularly spaced sources with a fixed delay, while in the present study, the limiting characteristic is spatially separated from the locus of new sources (see figures~\ref{Fig4} and \ref{Fig9} in the present paper). This aspect of the Euler equations means that the dynamics of the shock front influences the post-shock flow, which in turn influences the shock again, resulting in the entire flow being a nonsimple wave region. In the present study, the super-CJ velocity in a discrete reactive medium is a direct result of the significantly intense mechanical fluctuation near the limiting characteristic (as explained in \S~\ref{Sec_Analysis} and shown in figure~\ref{Fig9}), which is caused by the downstream propagating blast waves generated by the sources releasing energy. In contrast, due to the lack of left-running characteristics, a source releasing energy cannot generate a blast wave influencing the flow field downstream from the source in the Burgers equation system \citep{Mi2015}. Thus, we can hypothesize that it is the existence of both right- and left-running characteristics that is the necessary feature to observe the super-CJ solutions found in this study.

Extension of the current, Euler equation-based study to higher dimensions is likely to reveal richer behavior. Examination of the propagation of flames in three dimensional systems of random, discrete sources has revealed a percolation-like regime that has demonstrated the ability of flames to propagate beyond the thermodynamic limit of the corresponding homogeneous medium \citep{Tang2011PRE}, and similar behavior may occur with detonations in discrete systems. Examination of detonation propagation in the limit of spatially randomized, point-like sources in three dimensional clouds may have some relevance to the anomalous scaling between experiments in axisymmetric geometries (tubes) and two-dimensional slab geometries (rectangular geometries) \citep{Gois1996,Petel2006DS,Petel2007JL,Higgins2009APS,Higgins2013APS}.  A preliminary step in this direction was made in the recent computational study of \citet{Li2014CS}, in which a two-dimensional sinusoidal ripple in properties was introduced in a medium prior to detonation propagation through it. Extension of this study using perturbations with greater discreteness than sinusoidal would be of great interest.

Detonation theory has profitably explored the asymptotic limits of high activation energy \citep{BuckmasterNeves1988,Short1997highEa}, zero activation energy \citep{Erpenbeck1965}, high overdrive \citep{ClavinHe1996,ClavinHeWilliams1997}, low energy release \citep{ShortStewart1999}, and the Newtonian limit (i.e., ratio of specific heats approaching unity) \citep{Short1996Newtonian}. The line of investigation suggested in this study examines a different type of asymptotic limit, the limit of spatially discrete energy sources approaching $\delta$-functions in space and time. It is unknown if a rational asymptotic solution to the problem, rigorously derived from the governing Euler equations can be found, in contrast to the \textit{ad hoc} solution constructed in the Appendix. This problem is left as an open question to the detonation theory community.

\section{Conclusions}
\label{Sec_Conclusion}
Detonation propagation in systems with the energy release of the medium spatially concentrated into discrete pockets was simulated computationally.  The energy release of one such discrete source drives a blast wave, which is capable of initiating the next source after a prescribed delay.  The resulting ensemble of blast wave interactions propagates in a quasi-periodic manner, and the average wave speed was measured.  Systematic variation of the model parameters identified that the average wave deviated significantly from the CJ detonation speed of the equivalent homogenous media, with speeds as great as $15\%$ in excess of the CJ speed being measured.  This discrepancy is significant, given that experimental measurements of detonation speeds in gases usually agree to within $1\%$ of the equilibrium CJ speed.  A systematic variation of model parameters found that the deviation away from CJ depended on the degree of spatial concentration of the sources (more discrete sources resulted in greater deviations above CJ), with the wave speed reaching an asymptotic limit as the sources were concentrated into a space occupying less than $1\%$ of the entire domain (limit of $\Gamma \to 0$).  The deviation from CJ also depended on the ratio of specific heats (with greater deviations observed as $\gamma \to 1$), but the deviation from CJ did not depend upon the delay time of the sources between being shocked and releasing their energy or the average value of the energy release.  The results were interpreted via temporal averaging of the simulations onto a steady, one-dimensional projection that could be compared with the classical ZND structure of detonations.  This analysis suggests that the existence of non-CJ solutions can be interpreted as weak detonations due to the non-equilibrium of the flow resulting from the ensemble of shock interactions as the flow crosses the effective sonic surface.\\

The authors would like to acknowledge extended discussions with Samuel Goroshin, John Lee, Matei Radulescu, Vincent Tanguay, Hoi Dick Ng, and Charles Kiyanda that encouraged this study. Earlier simulations of the discrete detonation problem (the results of which, while not directly reported in this study, are consistent with the findings) were performed by A.J.H. using Clawpack (``Conservation Laws Package'') by R. Leveque. Additional simulations by Mehshan Javaid performed with E.T. provided further impetus and validation for this study.

\appendix
\section{Heuristic model}\label{appA}
\begin{figure}
\centerline{\includegraphics[width=1.0\textwidth]{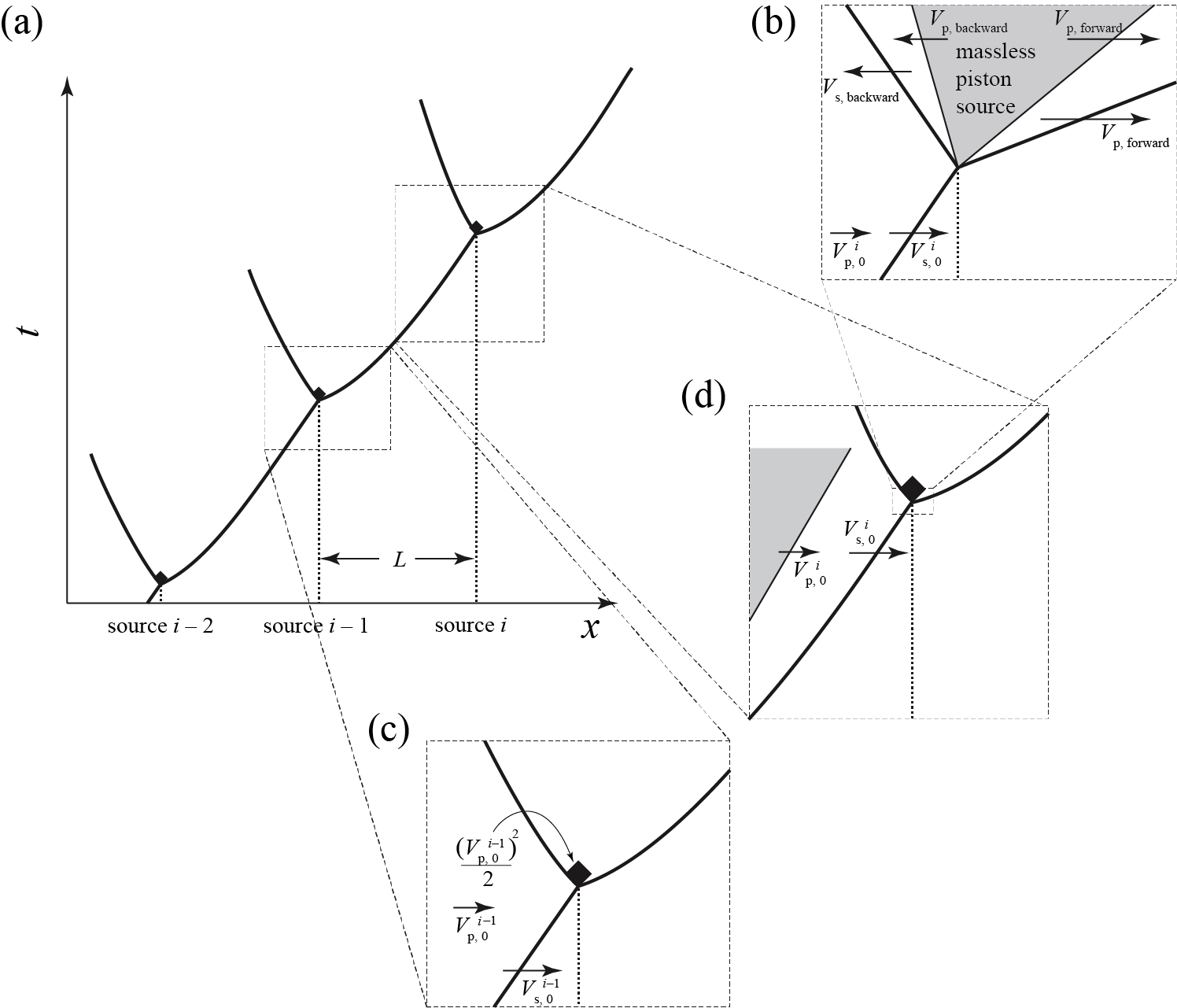}}
\caption{Schematic illustration of the one-dimensional, discrete-source detonation propagation problem, ($a$) as an $x$-$t$ (space-time) diagram. In ($b$), the mechanism of source energy deposition is assumed to occur via the impulsive motion of two outward-facing, massless pistons. The approximation of the carried-over influence from the previous sources onto the $(i-1)^\mathrm{th}$ source is shown in ($c$). In ($d$), the carried-over particle velocity imposed on the $i^\mathrm{th}$ source, $V_{\mathrm{p,}0}^{i}$, is approximated as the piston velocity required to sustain the instantaneous shock velocity, $V_{\mathrm{s,}0}^{i}$.}
	\label{Fig10}
\end{figure}

The collection of a large number of interacting blast waves and other unsteady flow features that comprise the dynamics of the discrete source detonation problem are unlikely to be amenable to an analytic solution.  In the limit of point-like energy sources (discreteness $\Gamma \to 0$) and the delay time going to zero ($\tau \to 0$), however, the problem becomes simpler.  In this case, as shown in figure~\ref{Fig10}($a$), the blast wave from a new source occurs on the shock front of the prior blast, initially with no other flow interactions involved.  This problem, and the subsequent propagation of the blast, are considered in this Appendix in an attempt to construct an analytic solution to this problem.  We emphasize that this is a constructed solution, in which priorly known solutions are patched together in an \textit{ad hoc}, or heuristic, manner, rather than a solution to the discrete detonation problem derived rigorously from the governing conservation laws.

An approximate, analytical solution to this problem for the case where the delay time and discreteness are both taken to the limit of zero begins with the similarity solution for the planar version of the well-known point-blast problem of Taylor and Sedov \citep{Jones1961}, which was inspired by the heuristic model of a detonation cell developed by \citet{Vasiliev1978} via a construction based upon the point-source blast wave solution. In this current model, the motion of the shock front $x_\mathrm{s}$ is given by the following non-dimensionalized equation,
\begin{equation}
\frac{\mathrm{d} x_\mathrm{s}}{\mathrm{d} t}=\sqrt{\frac{Q}{\mathrm{B}x_\mathrm{s}}}
\label{EqA1}
\end{equation}
where $\mathrm{B}$ is a dimensionless energy parameter depending on the specific heat capacity ratio $\gamma$.

Since the source releases its energy at the shock front originating from the previous source, the partitioning of blast energy into forward and backward propagating blast waves can be estimated as follows.  As shown in figure~\ref{Fig10}($b$), if the mechanism of source energy deposition is assumed to be two hypothetical massless pistons, one that pushes outward into the undisturbed gas ahead of the blast and the other that pushes into the gas behind the blast from the prior source, the pressure on both piston faces must be equal (since, being massless, they can exert no net force on the flow).  This condition permits the ratio of piston velocities to be solved for and, since the pistons must act for the same duration of time to conserve momentum, this ratio also determines the ratio of work done. Via the Rankine-Hugoniot normal shock relations, the ratio of particle velocity in front of each piston (i.e., piston velocity) and the work done by each piston can be obtained, and thus the energy partitioned to contribute to the foward- and backward-running blast waves can be determined. In the limit of strong shock relations, the equal pressure condition permits the partitioning of energy into forward and backward propagating blasts to be expressed analytically as a function of $\gamma$
\begin{equation}
\frac{Q_\mathrm{forward}}{Q}=\eta=\frac{1}{\sqrt{\frac{\gamma-1}{\gamma+1}}+1}
\label{EqA2}
\end{equation}
Since only $\eta Q$ contributes to the forward propagating blast, the equation describing the motion of the leading shock front needs to be modified
\begin{equation}
\frac{\mathrm{d} x_\mathrm{s}}{\mathrm{d} t}=\sqrt{\frac{\eta Q}{\mathrm{B}x_\mathrm{s}}}
\label{EqA3}
\end{equation}
Interestingly, (\ref{EqA2}) is the same partition of blast energy found by \citet{Sakurai1974} in examining planar blast waves generated by energy release at a stationary density interface (as opposed to energy release at a moving shock front in the current problem).

Following the release of source energy, the subsequent blast wave motion is additionally influenced by the particle velocity that was imposed by the blast wave from the previous source.  This problem (in position-time space) is analogous via hypersonic similarity to the two-dimensional hypersonic flow past a blunted wedge, a problem previously treated by \citet[pp. 218-221]{Chernyi1961}.  The blunt leading edge of the wedge represents the instantaneous energy release of the source under the hypersonic blast wave analogy, and the surface of the wedge represents the particle motion imposed by the previous blast wave (as determined by the blast strength from the previous source as it reached the new source). Considering the energy partitioned into the forward propagating blast and the particle motion imposed by the previous source, the motion of the blast wave propagation from one source to the next ($x_\mathrm{s}$ from $0$ to $1$) is given by
\begin{equation}
\frac{\mathrm{d} x_\mathrm{s}}{\mathrm{d} t}=\sqrt{\frac{\eta Q}{\mathrm{B} x_\mathrm{s}}}+V_\mathrm{p}\left(x_\mathrm{s}\right)
\label{EqA4}
\end{equation}
where $V_\mathrm{p}$ is the particle velocity carried over from the previous source. 

The carried-over particle velocity at the location of a newly triggered (the $i^\mathrm{th}$) source, i.e., $V_\mathrm{p}^{i}(x_\mathrm{s}=0)$ or $V_{\mathrm{p,}0}^{i}$, can be approximated as follows. The blast wave, which imposes $V_{\mathrm{p,}0}^{i}$ onto the $i^\mathrm{th}$ source, is a result of the energy release of not only the $(i-1)^\mathrm{th}$ source, but also all the previous sources (from the first one) with diminishing effect. The influence of the previous sources on the leading blast propagation after the trigger of the $(i-1)^\mathrm{th}$ source is, however, too complicated to be solved exactly. As shown in figure~\ref{Fig10}($c$), this influence is approximated as the kinetic energy possessed by the particle motion imposed on the $(i-1)^\mathrm{th}$ source, i.e., $\left(V_{\mathrm{p,}0}^{i-1}\right)^{2}/2$, added to the forward-partitioned source energy $\eta Q$ to propel the blast propagating towards the  $i^\mathrm{th}$ source. With this carried-over amount of energy from the previous sources, the velocity, at which the blast reaches the $i^\mathrm{th}$ source, can be obtained from the Taylor-Sedov solution
\begin{equation}
V_{\mathrm{s,}0}^{i} = \sqrt{\frac{\eta Q+\left(V_{\mathrm{p,}0}^{i-1}\right)^{2}/2}{\mathrm{B}}} 
\label{EqA5}
\end{equation}
The particle velocity imposed on the $i^\mathrm{th}$ source, $V_{\mathrm{p,}0}^{i}$, as shown in figure~\ref{Fig10}($d$), can be then approximated as the piston velocity required to sustain the blast front moving at its approximated instantaneous velocity $V_{\mathrm{s,}0}^{i}$
\begin{equation}
V_{\mathrm{p,}0}^{i} = \frac{2}{\gamma+1} V_{\mathrm{s,}0}^{i} = \frac{2}{\gamma+1} \sqrt{\frac{\eta Q+\left(V_{\mathrm{p,}0}^{i-1}\right)^{2}/2}{\mathrm{B}}} 
\label{EqA6}
\end{equation}
As the carried-over particle velocity is zero when the first source releases energy, the subsequent values of $V_{\mathrm{p,}0}$ can be calculated recursively. As the blast wave propagates to the next source, the influence of the previous source on the particle motion should diminish. In order to consider this effect, $V_\mathrm{p}$ is modeled to be inversely proportional to $x_\mathrm{s}$,
\begin{equation}
V_\mathrm{p}\left(x_\mathrm{s}\right) = \frac{1}{x_\mathrm{s}+1} V_{\mathrm{p,}0}
\label{EqA7}
\end{equation}
Note that the ``$1$'' in (\ref{EqA7}) is the spacing between two adjacent sources in the non-dimensionalized coordinates. This simple approximation can be justified be the fact that the particle velocity profile of a blast wave is roughly linear. Thus, the approximate, analytic solution of the leading shock wave propagation from one discrete source to the next can be obtained by integrating (\ref{EqA4}) with (\ref{EqA6}) and (\ref{EqA7}).

The predictions of this model for the instantaneous velocity of the shock front are shown in figure~\ref{Fig5}($a$).  Note that, since the Taylor-Sedov similarity solution for a point blast (in planar geometry) is used, the shock velocity at each source is initially infinite.  The velocity decays as the blast propagates forward, then jumps up again as a new source is triggered.  The minimum velocity before each new source is triggered monotonically increases as more sources are encountered, in qualitative agreement with the computational simulations.  In figure~\ref{Fig5}($b$), the average velocity (from one source to the next) is plotted as a dashed line, and again exhibits good qualitative agreement with simulations in the limit as discreteness $\Gamma \to 0$, although it tends to over-predict the final value of the plateau velocity.  This model can be used to examine the effect of $\gamma$, which enters the model significantly via the partitioning of blast energy released at the shock front (\ref{EqA2}), as shown in figure~\ref{Fig6}($b$).  Again, the model captures the qualitative trend of an increasing deviation away from the CJ solution as $\gamma$ approaches unity, although the predicted deviation is more than twice that observed in the simulations.

\bibliographystyle{jfm}
\bibliography{detonation}

\end{document}